\documentclass[11pt]{JHEP3}

\pdfoutput=1

\usepackage{amsmath}
\usepackage{amsfonts}
\usepackage{amssymb}
\usepackage{subeqnarray}
\usepackage{latexsym}
\usepackage[latin1]{inputenc}
\usepackage{xcolor}

\usepackage{epsfig}
\usepackage{graphicx}

\allowdisplaybreaks

\newcommand{\be}{\begin{equation}}
\newcommand{\ee}{\end{equation}}
\newcommand{\bea}{\begin{eqnarray}}
\newcommand{\eea}{\end{eqnarray}}
\newcommand{\bsea} {\begin{subeqnarray}}
\newcommand{\esea} {\end{subeqnarray}}
\newcommand{\al}{\alpha}
\newcommand{\ga}{\gamma}

\newcommand{\ba}{\beta}
\newcommand{\da}{\delta}
\newcommand{\ka}{\kappa}
\newcommand{\w}{\omega}
\newcommand{\Pp}{\mathbb P}
\newcommand{\Q}{\mathbb Q}
\newcommand{\ud}{\mathrm{d}}
\newcommand{\la}{\lambda}
\newcommand{\La}{\Lambda}
\newcommand{\eps}{\epsilon}

\newcommand{\G}{\mathcal G}
\newcommand{\F}{\mathcal F}

\newcommand{\ha}{\hat \alpha}
\newcommand{\ta}{\tilde \alpha}
\newcommand{\tar}{\tilde \alpha_r}
\newcommand{\mf}{\mathfrak m}
\newcommand{\qf}{\mathfrak q}
\newcommand{\V}{\mathcal V}

\renewcommand{\l}[1]{\left#1}
\renewcommand{\r}[1]{\right#1}

\newcommand{\cotanh}{ \mathrm{cotanh}}
\newcommand{\half}{\frac{1}{2}}
\newcommand{\e}{\mathrm{e}}
\def\nn{\nonumber}

\title{Higher-derivative scalar-vector-tensor theories: black holes, Galileons, singularity cloaking and holography}
\author{C. Charmousis$^{a,b}$, B. Gout\'eraux$^{c,d}$ and E. Kiritsis$^{c,d}$~\\
~\\
$^a$ \href{http://www.u-psud.fr}{Univ. Paris-Sud}, \href{http://www.th.u-psud.fr}{Laboratoire de Physique Th\'eorique}, CNRS UMR 8627, F-91405 Orsay, France;
~\\
$^b$ \href{http://www.phys.univ-tours.fr/}{LMPT}, Parc de Grandmont, Universit\'e Francois Rabelais, CNRS UMR 6083, Tours, France;
~\\
$^c$ \href{http://www.apc.univ-paris7.fr/APC_CS/}{APC}, AstroParticule et Cosmologie, Universit\'e Paris Diderot, CNRS/IN2P3, CEA/Irfu, Observatoire de Paris, Sorbonne Paris Cit\'e, 10 rue Alice Domon et L\'eonie Duquet, 75205 Paris Cedex 13, France;
~\\
$^d$ \href{http://hep.physics.uoc.gr}{Crete Center for Theoretical Physics}, Department of Physics, University of Crete, 71003 Heraklion, Greece.\\\strut\\
Email: \email{christos.charmousis@th.u-psud.fr}; \email{blaise.gouteraux@apc.univ-paris7.fr}; \href{http://hep.physics.uoc.gr/~kiritsis/}{http://hep.physics.uoc.gr/~kiritsis/}.
}

\preprint{CCTP-2012-13}

\abstract{We consider a general Kaluza-Klein reduction of a truncated Lovelock theory. We find necessary geometric conditions for the reduction to be consistent. The resulting lower-dimensional theory is a higher derivative scalar-tensor theory, depends on a single real parameter and yields second-order field equations. Due to the presence of higher-derivative terms, the theory has multiple applications in modifications of Einstein gravity (Galileon/Horndesky theory) and holography (Einstein-Maxwell-Dilaton theories). We find and analyze charged black hole solutions with planar or curved horizons, both in the 'Einstein' and 'Galileon' frame, with or without cosmological constant. Naked singularities are dressed by a geometric event horizon originating from the higher-derivative terms. The near-horizon region of the near-extremal black hole is unaffected by the presence of the higher derivatives, whether scale invariant or hyperscaling violating. For negative cosmological constant and planar horizons, thermodynamics and first-order hydrodynamics are derived: the shear viscosity to entropy density ratio does not depend on temperature, as expected from the higher-dimensional scale invariance.}

\keywords{Higher derivatives, Lovelock theory, Galileons, Holography}

\begin{document}

%\nocite{*}

\section*{Introduction}

Gravity theories admitting a finite number of degrees of freedom typically break down at very high curvature scales. This holds for General Relativity (GR), characterised by a unique metric degree of freedom, as well as  its obvious modifications  (such as Brans-Dicke gravity boosted with an extra massless scalar, \cite{Brans:1961sx}), up to richer gravity theories emerging from top-down approaches, as for example effective actions of string theories, \cite{Gross:1986mwMetsaev:1987zx}. Such {\it{effective}} theories of gravity can include a finite number of  tensor, vector or scalar  fields as well as  higher derivatives of these fields.

Higher-derivative actions appeared initially from interest in the higher energy behavior of gravitation and cosmology or as leading order corrections of string theories, a step beyond their point particle approximation, \cite{Gross:1986mwMetsaev:1987zx},\cite{Antoniadis:1993jc}. Higher-derivative interactions are also relevant in holographic applications when we  move away from the infinite 't Hooft coupling regime, see for example \cite{Brigante:2007nu}. Such higher-derivative extensions have a limited range of validity as they generically contain spurious solutions around the vacuum, \cite{Woodard:2006nt}. There are however rather special higher-derivative extensions of gravitational theories that have second-order field equations and at least have no problems (\emph{a priori})  with spurious solutions, \cite{Lovelock:1971yv}-\cite{Deffayet:2009mn}. Such well-defined higher derivative effective gravity theories, which we will call EGTs, will be the focal point in this article.

Although EGTs cannot appear as effective theories of a UV complete theory of gravity (which would produce an infinite number of terms and derivatives in the effective action), they are very useful laboratories in order to study higher-derivative effects in an analytically controllable setting. Furthermore, from a top-down approach they can provide a well-defined theoretical filter of classical alternatives to GR. They are therefore a tool which can be helpful in the above contexts.

In gravity and cosmology, attention focussed on EGTs in relation to the observed phase of late acceleration of the Universe. In an attempt to avoid introduction of a dark component of matter, GR may be modified at large distances deep in the IR (for a comprehensive review see \cite{Clifton:2011jh}). To that effect, scalar-tensor theories provide a simple yet non-trivial approach, being consistent limits of more complex theories of modification of gravity, whether higher-dimensional (such as DGP, \cite{Dvali:2000hr}), or four-dimensional (such as $f(R)$, which in fact is just a scalar-tensor theory in disguise, see \cite{Sotiriou:2008rpDeFelice:2010aj} for reviews).

The difficulty lies in crafting a model which successfully passes both solar system tests as well as strong field measurements from binary pulsars. A number of mechanisms have been devised in order to hide the effects of the scalar sector at small enough scales, such as higher-derivative effects, \cite{Amendola:2008vd}, the Chameleon, \cite{Khoury:2003aqKhoury:2003rn}, or the Vainshtein mechanisms, \cite{Deffayet:2001uk}.  Higher-derivative scalar-tensor versions of EGTs have recently been investigated under the name of Galileon theories. It is interesting to note that they originated from the decoupling limit of DGP gravity (where the extra scalar mode decouples from the other graviton polarisations), \cite{Nicolis:2008in}. The scalar Lagrangian is invariant under 'Galilean boosts' of the scalar field, hence the name. Suitably coupled to gravity, \cite{Deffayet:2009wt,Deffayet:2009mn,Deffayet:2011gz}, these symmetries only allow for a certain set of interaction terms, and have the nice property of maintaining second-order equations of motion. As it turns out, the most general scalar-tensor EGT in four dimensions was discovered independently long ago by Horndeski, \cite{Horndeski1974}. Furthermore, individual Galileon terms are known to arise from Kaluza-Klein (KK) reduction of the most general tensor EGT theory, namely Lovelock theory.
In accord to our EGT definition, Lovelock theory is a metric theory of gravity that preserves second-order equations of motion, \cite{Lovelock:1971yv} (see \cite{Charmousis:2008kc} for a review and references). It is made up of a finite series of dimensionally-extended Euler densities, which depend on the dimension of spacetime. In four dimensions, Lovelock theory reduces to GR. It is not too surprising therefore that Kaluza-Klein reductions of this theory produce a scalar(-vector)-tensor EGT, which is a part  of a Galileon theory.
Work in this direction was initiated by Horndeski, M\"uller-Hoissen and Kerner \cite{Horndeski:1978ca}-\cite{MullerHoissen1990709} giving a higher order  scalar-tensor and scalar-vector EGT. It was later-on pursued in a braneworld and cosmological context, \cite{Charmousis:2002rc}-\cite{GBcosmo},\cite{Amendola:2008vd}, while most recently in the Galileon picture, \cite{VanAcoleyen:2011mj}. Furthermore, scalar-tensor interacting Galileons have been shown to have interesting self-tuning properties, \cite{Fab4}, bringing a novel approach to treating the cosmological constant problem, \cite{Weinberg:1988cp}.

In the context of holography, applications to Condensed Matter systems were triggered by the study of the condensation of a complex scalar field in the vicinity of a charged black hole horizon, which was postulated to lie in the same universality class as high critical temperature superconductors (see for example the comprehensive reviews \cite{ReviewsHoloSC} and references within). More intriguing analogies were encountered studying holographically other Condensed Matter systems such as graphene or heavy fermion metals (see \cite{ReviewsHoloSC,AdSCMT,Sachdev:2012dq} for reviews), where no precise understanding of thermal critical phenomena or transport properties is available.

A common feature of these systems is their departure from traditional Fermi liquid behaviour in their normal - non-superconducting - phase, which signals a failure of the weakly-coupled description of the Fermi surface excitations in terms of quasi-particles, and is believed to originate from some quantum critical point, lying at zero temperature and hidden by the superconducting dome of the phase diagram. This quantum critical point might be (anisotropically) scale invariant, or even display hyperscaling violation - that is, the entropy scales with the temperature in some effective spatial dimensionality, \cite{Gouteraux:2011ce},\cite{Sachdev:2012dq}:
\be \label{hyperscaling}
	\mathfrak S\sim \mathfrak T^{\frac{d_{eff}}z}\,,
\ee
where $z$ is the dynamical critical exponent.\footnote{See \cite{kkp} for a review of recent experimental data and a change of paradigm implied by holographic models.}

Effective Holographic Theories (EHTs) are a useful tool to describe these unconventional matter phases. They are the application to holography of EGTs, and play the same role at strong coupling as standard effective Quantum Field Theories, \cite{Charmousis:2010zz,Meyer:2011xnGouteraux:2011xr}. Such theories always contain a metric (dual to the stress-energy tensor), a non-trivial field in any saddle-point solution. In principle, they contain an infinite number of fields as the dual QFTs contains an infinite number of single trace operators (that make up the spectrum of a string theory). However only a few of them (those with low dimensions in QFT or small masses in string theory) are important in the structure of the vacuum. Truncating the holographic theory to those important fields gives rise to the EHTs. Moreover, at very strong coupling one can neglect the higher-derivative terms in the gravitational theory.

 A simple set of EHTs are Einstein Maxwell Dilaton theories:
\be
	\label{ActionEMD}
	S_{EMD}=\frac1{16\pi G_N}\int\ud^{D}x\,\sqrt{-g}\,\left[R-\half\partial\phi^2-\frac14Z(\phi)F^2-V(\phi)\right].
\ee
The gauge field is dual to a conserved global current in the QFT, while the scalar drives the low energy dynamics and is dual to the most important scalar operator of the theory, \cite{Charmousis:2010zz,kachru,EMDHolo}. An important ingredient in the analysis of Einstein Maxwell Dilaton theories is the analytical study of asymptotics of physical observables, by a convenient parametrization of the two-derivative effective action, motivated by string-related supergravities, \cite{Charmousis:2010zz}:
\be \label{ExpCouplings}
	Z(\phi)\sim e^{\ga\phi}\,,\qquad V(\phi)\sim e^{-\da\phi}\,.
\ee
This class of theories was shown to generate the most general quantum critical (scaling) behavior for finite density systems, \cite{Gouteraux:2011ce,Huijse:2011efDong:2012se}, displaying (anisotropic) scale invariance or hyperscaling violation depending on the values of $\ga$ and $\da$. Therefore, it forms an important template for the low energy universality classes at finite density that can provide building blocks for more general systems along the lines proposed in \cite{jos}. In particular, $d_{eff}$  in \eqref{hyperscaling} was showed in \cite{Gouteraux:2011ce} to be directly related to 'hidden' scale invariance of some fictitious higher-dimensional theory, related to \eqref{ActionEMD}, \eqref{ExpCouplings} by generalised dimensional reduction (see below).

However,  the EHTs used so far included only up to two derivatives. To study further the dynamics when couplings (such as the 't Hooft coupling) are large but finite, higher-derivative interactions must be included. Indeed, these will become important close to the singularity for the electrically charged solutions as the scalar runs logarithmically and the string coupling diverges, \cite{Chan:1995fr, kachru} (in the magnetic case on the other hand, quantum effects will need to be taken into account, \cite{EMDmagnetic}). Moreover, transport coefficients such as the shear viscosity, which control the out-of-equilibrium behaviour and obey universal laws in two-derivative setups, \cite{Kovtun:2004de}, may violate these laws upon turning on higher derivatives, \cite{Brigante:2007nu}. It is therefore worthwhile to study how these departures from universality occur, as well as their dependence on temperature and density.

A central role in EGTs is played by their exact solutions. Amongst them, their black hole solutions stand out as strongly gravitating backgrounds protected in their UV sector by the presence of an event horizon. They are essential for the understanding of screening mechanisms, like Vainshtein's, \cite{Bab}, and strong gravity astrophysical phenomena. Black holes are furthermore thermal backgrounds with calculable thermodynamic or transport quantities. Beyond General Relativity, specific EGTs can admit black holes with primary or secondary hair, permitting non-trivial fields to be switched on in their horizon vicinity, and allowing a richer phase structure of solutions for given asymptotic conditions (see for example \cite{HairyBHs}). They provide in this sense multiple, regular saddle points in specific thermal baths presenting often phase transitions from non-hairy to hairy black holes as the temperature of the heat bath is lowered.

Exact black hole solutions have been known for a long time both in Lovelock theory, \cite{Boulware:1985wk}-\cite{Myers:1988ze}, and for two-derivative scalar-tensor theories, \cite{Chan:1995fr}-\cite{Charmousis:2009xr}. In the latter, dilatonic black holes can display unusual asymptotics, departing from AdS, but still allow for a consistent description. However, they have proven hard to find analytically with higher-derivatives, and only perturbative or numerical results are known  \cite{Horndeski:1978ca,Mignemi:1988qcWiltshire:1990ah,GBDsols}. Motivated by the diverse considerations above, we would like to make progress on the front of exact black hole solutions in higher-derivative theories. As we shall see, the key in obtaining these combines two ingredients: Lovelock theory and generalised (Kaluza-Klein) dimensional reduction, \cite{Kanitscheider:2009as,Gouteraux:2011ce,Gouteraux:2011qh}.

A dimensional reduction is called \emph{generalised} if it is a consistent reduction (every solution of the lower-dimensional field equations can be lifted to a solution of the higher-dimensional field equations) and the number of reduced dimensions is kept arbitrary and can be analytically continued in the lower-dimensional theory to a \emph{continuous, real parameter}. Then, all properties of the usually complicated, lower-dimensional setup can be inferred \emph{via} the reduction. For instance, \cite{Kanitscheider:2009as} derived the first-order hydrodynamic transport coefficients of non-conformal branes, \cite{Boonstra:1998mp}, by connecting them to AdS black branes. In \cite{Gouteraux:2011ce}, it was found that the scaling solutions of Einstein Maxwell Dilaton theories studied in \cite{Charmousis:2010zz} with running dilaton can be uplifted to AdS or Lifshitz solutions of a higher-dimensional gravitational EHT. This explains their scaling behavior as well as the special values of the dilaton functions that separate gapless from gapped solutions. In \cite{Gouteraux:2011qh}, the effect of non-diagonal KK vectors was investigated.

To sum up, our goal in this work is to make progress on the front of the analytical study of higher-derivative scalar(-vector)-tensor theories, by starting from a controlled setup (Lovelock theory) where ghosts are absent, and using generalised dimensional reduction, which is essential to our analytical treatment. As advertised above, the solutions we obtain are relevant both for modified gravity theories (Galileons) - providing the first spherically symmetric analytical black hole solutions in this setup - and for holography - in particular for determining the fate of the near-horizon scaling geometries under higher-derivative corrections.

The plan of the rest of paper is as follows. In section \ref{Section:CurvedDiagKKRed}, we perform a \emph{generalised} diagonal reduction from an arbitrary-dimensional Einstein Gauss-Bonnet theory and show that it is consistent if the compactified space is a Gauss-Bonnet space, which generalises the usual notion of an Einstein space. The result is a scalar-tensor theory with quartic derivatives in the metric and scalar field, depending on a continuous, real parameter. The full details of the reduction are enclosed in Appendix \ref{App:DiagonalKKRed}. In section \ref{Section:EGBD},  we take advantage of our reduction scheme to derive a family of static, spherically symmetric black hole solutions, with arbitrary horizon topology. To the best of our knowledge, these are the first exact analytical black hole solutions derived in a scalar-tensor theory with higher-derivative interactions. We study successively the planar and spherical cases, and then turn to the Galileon frame. In section \ref{Section:HoloEGBD}, we describe how to set up the holographic dictionary for toroidal reductions, and then derive the thermodynamics and first-order hydrodynamic transport coefficients for the solutions of section \ref{Section:EGBD}. Finally, in section \ref{Section:S1NonDiagKKRed}, we turn to a non-diagonal reduction on a circle, which cannot be generalised, derive an exact black hole solution, sketch how to set up holography for this class of theories, and study thermodynamics and first-order hydrodynamics. Section \ref{Section:CCL} contains our conclusions.

\section{Generalised Kaluza-Klein reduction of Einstein Maxwell Gauss-Bonnet theories\label{Section:CurvedDiagKKRed}}

Our starting point is the $D$-dimensional Einstein Gauss-Bonnet action with a $U(1)$ Maxwell field. This term represents the 5 or 6-dimensional Lovelock theory; although we consider $D$ to be arbitrary, we will truncate higher-derivative terms at this level for simplicity. We have
\be
	S = \frac1{16\pi G_N}\int \ud^{D}x\,\sqrt{-g}\l[-2\La +R+\ha\G-\frac{1}4F^2\r],
\label{GBAction}
\ee
where
\be
	\G=R_{ABCD}^2-4R_{AB}^2+R^2
	\label{GB}
\ee
is the Gauss-Bonnet curvature invariant. Metric variation of the action gives
\be
	\mathcal E_{AB} = G_{AB}+\La g_{AB} -\ha H_{AB}=\half F_A^CF_{BC}-\frac18F^2g_{AB}
	\label{GBEOM}
\ee
with $G_{AB}$ the Einstein tensor and $H_{AB}$ the Lanczos tensor
\be
	H_{AB}  = \frac{{g_{AB} }}{2}\G - 2RR_{AB}  + 4R_{AC} R^C _{\;B}  + 4R_{CD} R^{C\;D} _{\;A\;B}  - 2R_{ACDE} R_B ^{\;CDE} \,.
	\label{LanczosTensor}
\ee

We consider a diagonal reduction along some arbitrary $n$-dimensional internal space $\tilde{\mathbf  K}$, reducing down to $p+1$ spacetime dimensions ($D=p+n+1$):
\be
	\ud s^2 = e^{-\da\phi}\ud \bar s^2 +  e^{\frac{\phi}{\da}\l(\frac2{p-1}-\da^2\r)}\ud \tilde K^2\,,\qquad \frac{p-1}2\da^2 = \frac{n}{n+p-1}\,. \label{KKStatic2}
\ee
All terms with a tilde will refer to the $n$-dimensional internal space, while terms with a bar will refer to the $(p+1)$-dimensional theory. By exchanging $n$ - the arbitrary integer number of reduced dimensions - for $\da$ and analytically continuing the latter to the whole real line, this reduction is generalised in the manner of \cite{Kanitscheider:2009as,Gouteraux:2011qh}. This analytic continuation is possible if and only if one shows the consistency of the KK reduction for an arbitrary number of dimensions, \emph{i.e.} the reduced equations of motion are derived from the reduced action. In Appendix \ref{App:DiagonalKKRed}, we give the full details of the reduction and show that it is consistent if the internal space $\tilde{\mathbf K}$ is a Gauss-Bonnet space. A Gauss-Bonnet space is characterised by two properties: it is an Einstein space whose Lanczos tensor \eqref{LanczosTensor} is also proportional to the metric, \eqref{GBSpace}.
At the end of the day the reduced action reads,
\be \label{CurvedKKGBAction}
\begin{split}
	\bar S =& \int \ud^{p+1}x\,\sqrt{-\bar g}\l\{\bar R\l[1+2\ha \tilde Re^{-\frac{\phi}{(p-1)\da}\l(2-(p-1)\da^2\r)}\r] -2\La e^{-\da\phi}+\tilde R e^{-\frac{2\phi}{(p-1)\da}}-\frac{e^{\da\phi}}4F^{\;2}\r.\\
			&+\ha\tilde\G e^{-\frac{\phi}{(p-1)\da}(4-(p-1)\da^2)}-\half\partial\phi^2\l[1+2\ha c_5(\da,p)\tilde Re^{-\frac{\phi}{(p-1)\da}\l(2-(p-1)\da^2\r)}\r]\\
			&\left.+\ha e^{\da\phi}\l[\bar \G + c_2(\da,p)\bar G^{\mu\nu}\partial_\mu\phi\partial_\nu\phi +c_3(\da,p)\l(\partial\phi^2\r)^2+ c_4(\da,p)\partial\phi^2\square\phi  \r]\r\}.
\end{split}
\ee
Note the higher order Gauss-Bonnet coupling $\ha$ that parametrises the presence of higher order corrections emanating from the Gauss-Bonnet invariant.
The $c_i(\da,p)$ are coefficients fixed by the Kaluza-Klein reduction:
\bea
	c_2(\da,p)&=&\frac{2[n(p-3)+(p-1)^2]}{(p-1)(n+p-1)}=2(1-\da^2)\,,\\
	 c_3(\da,p)&=&\frac{2(p-1)^2-n(p^2-1)-n^2(p-3)}{4n(p-1)(n+p-1)}=\frac{3\da^2}4-\frac{p+5}{4(p-1)}+\frac{\da^{-2}}{(p-1)^2}\,,\\
	c_4(\da,p)&=&\frac{\sqrt{2}(n+1-p)}{\sqrt{n(p-1)(n+p-1)}}=\frac{2}{(p-1)\da}\l[(p-1)\da^2-1\r],\\
	c_5(\da,p)&=&\frac{n(p+n-5)-6(p-1)}{n(n+p-1)}=-\da^2+\frac{p+7}{p-1}-\frac{12}{(p-1)^2\da^2}\,.
\eea
The real number $\da$ parametrises the theory according to the above coefficients.
Generically we see that the reduction has given us a triple exponential potential, namely
\be
\label{effpot}
V_{eff}=-2\La e^{-\da\phi}+\tilde R e^{-\frac{2\phi}{(p-1)\da}}+\ha\tilde\G e^{-\frac{\phi}{(p-1)\da}(4-(p-1)\da^2)},
\ee
and a number of higher order kinetic terms including the $(p+1)$-dimensional Gauss-Bonnet density. Even for $p=3$ this term is no longer topological since it couples to the scalar field $\phi$.

When we switch off the higher order coupling constant $\ha=0$, we obtain an Einstein-Maxwell-dilaton theory with a double exponential potential, originating from the higher-dimensional cosmological constant and the Ricci curvature $\tilde R$ of the compactified space. Moreover, the frame is then the Einstein frame (for a nice discussion on parametrisations of scalar-tensor theories see \cite{EspositoFarese:2000ij}), which motivated our choice of the conformal factor in (\ref{KKStatic2}). Note that the action is symmetric under the exchange
\be \label{AdSFlatMap}
	-2\La \longleftrightarrow \tilde R\,, \qquad \da\longleftrightarrow \frac{2}{(p-1)\da}\,,
\ee
as was first noted in \cite{Charmousis:2003wm} in the case of Weyl geometries.
This symmetry in the action does not survive the introduction of a non-zero coupling, $\ha\neq0$.

The case where the compactified/internal space is not flat, \cite{Charmousis:2003ke}, \cite{Amendola:2005cr}, turns out to be far more complex since terms proportional to the internal Ricci and Gauss-Bonnet scalar appear: some are just additional exponential terms in the effective potential (\ref{effpot}) and do not present any particular difficulty, but two terms come and renormalise the kinetic term for the scalar field and the coefficient in front of the $(p+1)$-dimensional Ricci scalar. Thus, the lower-dimensional frame is not generically the Einstein frame. When we have a flat internal space, we could still characterise (\ref{CurvedKKGBAction}) as being the Einstein frame with the caution that the scalar degree of freedom is {\it still} not minimally coupled to the metric due to (any of) the higher order kinetic terms (see for example \cite{Deffayet:2010qz}). All these considerations put aside, for definiteness, this is the frame we will consider in the forthcoming sections having in mind holographic or stringy applications.

%If one insists on  having minimal coupling between gravity and the scalar field at zeroth order in $\ha$, one could transform conformally to the metric
%\be
%	\tilde g_{\mu\nu}=h(\phi)^{\frac2{p-1}} g_{\mu\nu}\,, \qquad h(\phi)=1+2\ha R_{(n)}e^{-\frac{\phi}{(p-1)\da}\l(2-(p-1)\da^2\r)}\,.\label{ConfTrans}
%\ee
%{\bf Blaise: I think we should impose a constraint such that the sign of $h$ should be positive, so that the sign in front of the Einstein term in the action remains the correct one. Though things may be different since we are not in the Einstein frame. Christos, what do you think?}

Another interesting frame to consider is the frame where there is no conformal factor of $\phi$ in front of the $(p+1)$-dimensional part of the reduction Ansatz:\footnote{
In holographic studies, this frame is called the \emph{dual} frame, where one can most easily set up the holographic dictionary for non-conformal branes, as they are asymptotically AdS in this particular frame, \cite{Boonstra:1998mp,Kanitscheider:2008kd,Kanitscheider:2009as,Gouteraux:2011qh}.
}
\be
	\ud s^2_{(p+n+1)}=\ud \bar s^2_{(p+1)} + \e^{\phi}\ud \tilde K^2_{(n)}
	\label{KKGalileon}\,.
\ee
In \eqref{KKStatic1}, this amounts to setting $\al=0$ and absorbing $\beta$ in $\phi$. Alternatively, one may also perform a conformal transformation in the action \eqref{CurvedKKGBAction}.
 In this frame, the reduced action reads
\be\label{CurvedGBGalileonAction}
\begin{split}
	\bar S_{galileon}=& \int \ud^{p+1}x\,\sqrt{-\bar g}\,e^{\frac n2\phi}\l\{\bar R -2\La +\ha\bar \G+\frac n4(n-1)\partial\phi^2 -\ha n(n-1)\bar G^{\mu\nu}\partial_\mu\phi\partial_\nu\phi \r.\\
			&-\frac\ha4n(n-1)(n-2)\partial\phi^2\square\phi+\frac\ha{16}n(n-1)^2(n-2)\l(\partial\phi^2\r)^2\\
			&\left.+e^{-\phi}\tilde R\l[1+\ha \bar R+\ha 4(n-2)(n-3)\partial\phi^2\r]+\ha\tilde\G e^{-2\phi}\r\}\,,
\end{split}
\ee
This action can be related to so-called Galileon actions, \cite{Nicolis:2008in,Deffayet:2009wt,Deffayet:2009mn,Deffayet:2011gz}: it has been shown in \cite{VanAcoleyen:2011mj} how covariant Galileon actions could be obtained by dimensionally reducing from Lovelock actions. In this work, we truncate to the second Lovelock order, that is up to the Gauss-Bonnet term.
Again the reduction is a consistent one, meaning that all lower-dimensional solutions can be uplifted to $p+n+1$ dimensions, and conversely that all higher-dimensional solutions of Einstein-Gauss-Bonnet theories respecting the symmetries of \eqref{KKGalileon} can be reduced to yield a solution of the second-order Galileon action. Finally, remember that the reduction is \emph{generalised}, which means that the number of reduced dimensions $n$ in \eqref{CurvedGBGalileonAction} can be analytically continued to the real axis and thus generates a continuous family of theories, labeled by $n$. Here, it is interesting to note that neglecting the Einstein-Hilbert term, in \eqref{GBAction} (or taking $\ha$ infinite), considering flat compactification and $n=1$ gives three out of the four 'Fab 4' terms \cite{Fab4} for specific exponential couplings.
In particular, notice that one picks up an ordinary Einstein-Hilbert term from reduction of the higher order Gauss-Bonnet term as has been noticed in codimension 2 braneworld scenarios, \cite{CoD2BW}.

The Galileon field can then simply be understood to be the scalar parameterising the volume of the internal space. Reducing from the Einstein - Gauss-Bonnet action yields all the terms up to quartic order in derivatives (either of the metric or the scalar, or a mixed combination of the two). Reducing higher order Lovelock densities will yield terms with a higher number of derivatives. A typical example is the Paul term appearing in Fab 4 theory \cite{Fab4} which involves six derivatives but still gives second order field equations. This term originates from the third order Lovelock density \cite{VanAcoleyen:2011mj,Fab4}. Reinterpreted in the higher-dimensional picture, it is quite natural why second-order equations of motion should derive from the Galileon actions. Finally, the shift symmetry of the scalar can now be construed as the lower-dimensional truncation of the higher-dimensional symmetries to those diffeomorphisms which leave the reduced metric invariant.

\section{Charged dilatonic Einstein Gauss-Bonnet black holes\label{Section:EGBD}}

Having the reduced higher order theories at hand, we shall review the seed black holes in $D$-dimensional Gauss-Bonnet theories (\ref{GBAction}) and their basic properties. We will then dimensionally reduce these in the two different frames using the reduction we established in section \ref{Section:CurvedDiagKKRed}.

\subsection{Black holes in Einstein Gauss-Bonnet theories \label{Section:GBBH}}

Start by defining the reduced Lovelock coefficients in spacetime dimension $D$
\be
	L^{-2}=\frac{-2\La}{(D-1)(D-2)}\,, \quad \ta = 2\ha (D-3)(D-4)\,. \label{ReducedLovelockCoeff}
\ee
The black hole solutions\footnote{Spherically symmetric black hole solutions with maximally symmetric horizon ($\Theta=0$) were first exhibited in \cite{Boulware:1985wk}-\cite{Wheeler:1985nh}, while topological AdS black holes were studied in \cite{Cai:2001dz}, and dS ones in \cite{Cai:2003gr}.} to action (\ref{GBAction}) take the form
\bea
	\ud s^2_{(D)} &=& -V(\rho)\ud t^2 + \frac{\ud \rho^2}{V(\rho)}+\rho^2\ud K^2_{(D-2)}\,, \label{BHMetric}\\
	V(\rho)&=& k+\frac{\rho^2}{\ta}\l[1\mp\sqrt{1-\frac{2\ta}{L^{2}}-\frac{\ta^{2}\Theta}{\rho^4}+\frac{4\ta \mf}{\rho^{D-1}}-\frac{2\ta\qf^2}{\rho^{2(D-2)}}}\r], \label{EGBBHPot}\\
	A&=&\sqrt{2\frac{(D-2)}{(D-3)}}\frac{\qf}{\rho_+^{D-3}}\l(1-\l(\frac{\rho_+}{\rho}\r)^{D-3}\r)\ud \tau\,,
\eea
where the '$-$', Einstein, and '$+$', Gauss-Bonnet branch are to be distinguished.\footnote{Note that the $+$ branch vacuum is perturbatively unstable, \cite{Charmousis:2008ce} and does not have a smooth $\ha \rightarrow 0$, Einstein limit. For these two reasons and for what follows, we shall focus on the $-$ branch.}

The IR AdS vacuum scale is renormalised to
\be
	\La_e = \frac{(D-1)(D-2)}{2\ta}\l(1-\sqrt{1+\frac{2\ta}{L^2}}\r) \label{AdSEffLambda}\,.
\ee
Though the coupling to Gauss-Bonnet renormalises the bare cosmological constant, it has no influence on the sign of the effective one: one may check from \eqref{AdSEffLambda} that $sign(\La_e)=sign(\La)$. Finally, whenever $\ha\La<0$, one must take care that the square root in \eqref{AdSEffLambda} is well-defined:
\bea
	\La<0 &\Rightarrow& 0\leq\ta\leq\ta_{max}\,,\qquad \ta_{max}=\frac{(D-1)(D-2)}{(-4\La)}\,,\\
	\La>0 &\Rightarrow& \ta_{min}\leq\ta\leq0\,,\qquad \ta_{min}=-\frac{(D-1)(D-2)}{4\La}\,.
\eea
The limiting case where these bounds are saturated is special: there exists a single vacuum to the theory, with an enhanced symmetry group. For instance, in odd dimensions, the theory can be rewritten in one dimension higher as a Chern-Simons theory, \cite{Zanelli:2005saCrisostomo:2000bb}.

%Note that, quite differently from the Einstein case, the \emph{geometry} of the horizon appears in the black hole potential through the (possibly) non-trivial $\Theta$ term.
The integration constants $\mf$ and $\qf$ are related to the mass and charge of the black hole. The constants $k$ and $\Theta$ are related to the geometry of the horizon sections, $\mathbf K^{D-2}$.
If we discard momentarily the $\Theta$ term appearing in the black hole potential, an expansion for large $\rho$ gives us the RN plus cosmological constant solution of higher-dimensional GR. However,
in order for \eqref{BHMetric} to be a solution for the higher order theory, $\mathbf K^{D-2}$ must be a \emph{Gauss-Bonnet} space, which is a space verifying both the usual Einstein space condition (as for GR)
\be
	R_{ij} = (D-3)k h_{ij}, \qquad  k=0,\pm1 \; \textrm{for a constant curvature space,}\label{EinsteinSpace}
\ee
but also that the Lanczos tensor of the horizon should also be proportional to the horizon metric. For an Einstein metric this translates to the following condition on the Weyl tensor
\be
\label{GBcond}
	C_{iklm}C^{jklm}=\frac{(D-3)!}{(D-6)!}\Theta\da^j_i\,,
\ee
where $\Theta$ is \emph{a constant}. This term first appeared as an obstruction to Einstein type horizons for Lovelock theory in \cite{Dotti:2005rc}, and later the black hole solutions were investigated with specific examples in $D=6$ in \cite{Bogdanos:2009pc} (see also \cite{Dotti:2010bw}).
%This identity on the Weyl tensor\footnote{See \cite{Edgar:2001vv} for other useful identities on the Weyl tensor.} is trivial for four-dimensional horizons, as it is simply a restating that the $n$-th Lovelock tensor vanishes identically in dimension $2n$. The upshot of the Gauss-Bonnet term in $D=6$ is that the square of the Weyl tensor is a constant.

In this work, we shall consider a specific horizon geometry which realises the Gauss-Bonnet space condition: a product of maximally symmetric spaces with equal radii (see  Appendix \ref{Appendix:GBNonMaxHor} and \cite{Maeda:2010bu} for proof). Setting $D-2=m(1+s)$ with $m$ and $s$ two positive integers, the horizon has geometry $\left(\mathbf K^m\right)^{1+s}$, where the $\mathbf K^m$ are maximally symmetric spaces with equal radii and curvature $\ka=\pm1,0$. Then, one shows that
\bea
	k&=&\frac{(m-1)}{D-3}\ka\,,\qquad m(1+s)=D-2\,,\qquad \ka=\pm1,0\,, \label{kDef}\\
	\G&=&(D-2)(m-1)\l[(D-2)(m-1)-2(2m-3)\r]\ka^2\,,\\
	\Theta&=&\frac{(D-6)!}{(D-2)!}\G - k^2 = \frac{2(D-m-2)}{(m-1)(D-4)(D-5)}\ka^2.
\eea
Setting $s=0$ and $m=D-2$ in the above recovers the usual formul\ae\ for homogeneous spaces (in particular $\Theta=0$).

This induces an important difference between GR and EGB theory. Take $D=6$ in order to fix the discussion and as our horizon consider the homogeneous space $\mathbf S^{4}$ ($m=4$, $s=0$): $k=+1$ from \eqref{kDef}. $\Theta$ is obviously zero and we are in the Boulware-Deser family of solutions, \cite{Boulware:1985wk}. Take now a non-homogeneous Einstein space $\mathbf S^{2}\times \mathbf S^{2}$ ($m=2$, $s=1$) with both spheres having the same curvature. In this case, it is easy to see that $\Theta$ is constant and non-zero. In GR, although the black hole potential \eqref{EGBBHPot} is identical for both of the horizon geometries, the normalised constant $k$ appearing in the latter case is not unity for each unit sphere, in fact we can see that $k=1/3$ from \eqref{kDef}. This not only changes the asymptotics but also renders the $\mf=0$ solution singular with a real curvature singularity at $\rho=0$. The situation is similar to the gravitational monopole,{\footnote{We thank Eugeny Babichev for discussions on this subject.}} \cite{Barriola:1989hx}, except that here there is no matter core to smooth out the singularity at $\rho=0$. Of course the singularity can be screened from a far away observer by having a large enough mass $\mf$. This is the slightly pathological situation in GR: a mass is needed to regularise the geometry. In Lovelock theory however, there is an important change, for the presence of the extra curvature term induces a contribution in the potential \eqref{EGBBHPot} which can  actually cover the singularity, \cite{Bogdanos:2009pc}, giving us a regular gravitational monopole-like solution. Thinking in effective theory terms, we have a precise example where the leading stringy correction can actually dress an otherwise naked singularity. We will come back to this when looking closer at four-dimensional dilatonic black holes in sections \ref{Section:CurvedEGBD} and \ref{section:GalBH}, where we will want to consider multiples of $m$-spheres in order to have a locally $\mathbf S^{m}$ horizon black hole.

Since the metric is static, zeros of \eqref{EGBBHPot} correspond to event horizons, $r=r_h$, while $r=0$ is the central curvature singularity, and wherever there is a zero of the square root we have a branch singularity at $r=r_s\geq 0$. The spacetime is a black hole if and only if $0\leq r_S<r_h$ and $f(r>r_h)>0$, notwithstanding the occurrence of a cosmological horizon.

\subsection{Planar dilatonic black hole\label{Section:PlanarEGBD}}

The simplest case is the toroidal reduction, which gives rise to a black hole with a planar horizon. This solution is a $0$-brane in $p+1$ dimensions and is of particular interest as a finite temperature background for holographic applications. The reduced action \eqref{CurvedKKGBAction} is simplified since all "tilded" geometric terms are identically zero given that the internal space is flat. This is a requirement from the fact that we want to have a planar horizon black hole in $p+1$ dimensions and hence the $D$-dimensional black hole \eqref{BHMetric} must also have a flat horizon given condition \eqref{EinsteinSpace}.{\footnote{There may be Euclidean signature geometric spaces such that we have a Ricci flat space but with $\Theta\neq 0$. This is an open question and would be an interesting extension to what we consider here.}} Obviously here we are interested in locally AdS (instead of flat or dS) type of asymptotics in order to have an event horizon with planar geometry in $D$ dimensions.
Comparing the KK Ansatz \eqref{KKStatic2} and \eqref{BHMetric}, we can identify the lower-dimensional fields (after some rescalings of the coordinates and integration constants):
\bea
	\ud s^2_{(p+1)} &=&r^2\l[-f(r)\ud t^2+\ud R^2_{(p-1)}\r]+\frac{r^{(p-1)\da^2-2}\ud r^2}{f(r)},\label{MetriciKKToroidal}\\
	f(r)&=&\frac{1}{\tar}\l[1\mp\sqrt{1-2\tar\l(\frac{1}{\ell^2}-\frac{2 \mf}{r^{p-\frac{p-1}2\da^2}}+\frac{\qf^2}{r^{2(p-1)}}\r)}\r],\label{BHPotKKToroidal}\\
	 e^{\phi}&=&r^{(p-1)\da}\,,\label{PhiKKToroidal}\\
	A&=&-\sqrt{\frac{2(p-1)}{(p-2)+\frac{(p-1)\da^2}{2}}} \frac{\qf}{r^{(p-2)\frac{(p-1)\da^2}{2}}}\ud t\,,\label{AKKToroidal}
\eea
having defined the lower-dimensional Lovelock coefficients
\be
	\ell^{-2}=\frac{-2\La}{(p-1)\l(p-\frac{p-1}2\da^2\r)}\,,\quad  \tar=2\ha\Big(p-2+\half(p-1)\da^2\Big)(p-3+(p-1)\da^2). \label{DilatonicReducedLovelockCoeff}
\ee
Note that
\be
\label{l}
sign(\tar)=sign(\ha)\,,\qquad  \ell\in\mathbf R\Rightarrow \La\l(p-\half(p-1)\da^2\r)<0\,.
\ee
and $\mf$, $\qf$ are respectively related to the mass and charge of the solution.
This planar black hole reduces to the usual $\ga=\da$ Einstein-Maxwell-Dilaton solution\footnote{See \cite{Mignemi:1988qcWiltshire:1990ah} for the neutral version, and \cite{Cai:1996eg}-\cite{Charmousis:2009xr},\cite{Charmousis:2010zz}, for the charged version of the dilatonic Einstein black hole with a planar horizon. See also \cite{Cai:2004iy} for an earlier study of the interpretation of Einstein dilatonic black holes \emph{via} dimensional reduction.} upon taking the $\tar\to0$ limit of the Einstein branch. In fact here, our effective potential \eqref{effpot} is just the usual Liouville potential $V_{eff}=-2\Lambda e^{-\delta \phi}$. The $\da=0$ limit correctly recovers the non-dilatonic Gauss-Bonnet planar black hole (with a note of caution for $p=3$ since $\tar \rightarrow 0$).
For values of $\da$ such that $(p-1)\da^2/2<p$, the mass term under the square root decays asymptotically. Clearly the asymptotes are unusual given the fall-off of the mass term, compared to the non-dilatonic case, $\da=0$.
Consider then  $\da^2<2p/(p-1)$. Since the black hole is static, any zeros of the potential \eqref{BHPotKKToroidal}
will be coordinate singularities (horizons): we can define Kruskal coordinates in the usual way,
\be
dv_{\pm}=dt\pm \frac{dr}{f(r)}
\ee
and the charts $(v_{\pm},r)$ are then regular in the past and future light cone. However, apart from the central singularity at $r=0$, we have to also make sure that there are no zeros of the square root in \eqref{BHPotKKToroidal}, which correspond to additional branch singularities due to the higher order corrections in $\tar$. Already our vacuum, $(\mf,\qf)=(0,0)$ is restricted by the condition,
\be
2\tar \leq \ell^2\,,
\ee
which when saturated corresponds to the strong coupling case \cite{Charmousis:2008ce} sometimes referred to as the Chern-Simons (Born-Infeld) limit in odd (even) dimensions in the vacuum higher-dimensional solutions (see also \cite{Zanelli:2005saCrisostomo:2000bb}). In this limit the two branches merge and the (in)stability of the vacuum cannot be decided in a linear perturbation manner, \cite{Charmousis:2008ce}. In loose terms, the closer we are to this saturation point, the bigger the effect of the higher order terms in $\tar$.

In order to analyse the horizon structure of the solution, we follow Myers and Simon, \cite{Myers:1988ze} (see also \cite{Charmousis:2008kc}). Define the polynomials $\Q$ and $\Pp$:
\bea
\Q(x)&=&x^{2(p-1)}-\tar \Pp(x) \label{Q1}\,,\\
\Pp(x)&=&\frac{2}{\ell^2}x^{2(p-1)}-4 \mf x^{p-2+\frac{p-1}{2}\da^2}+2 \qf^2\,.
\eea
Note that $\Pp$ is of the same sign as $f(r)$, so it is easy to determine the nature of the spacetime (black hole or cosmological) from the large $r$ behaviour of $\Pp$. The zeros of $\Pp$ determine horizons and are identical to the zeros of \eqref{BHPotKKToroidal} for $\tar=0$, hence horizon positions are independent of the coupling $\tar$. One can show that the zeros of $\Q$ are branch singularities, and given \eqref{Q1}, we have $\Q(r_h)>0$  wherever $\Pp(r_h)=0$. If $0<2\tar\leq \ell^2$, $\Q$ is an increasing function and then zeros of $\Pp$ are to the right of branch singularities. Thus,  for large enough mass compared to charge, we will have an inner and outer event horizon, exactly as for $\tar=0$, \cite{Charmousis:2009xr,Charmousis:2010zz}. When $\tar <0$ and small enough, the inner and event horizon remain intact. However, increasing the magnitude of $\tar$, the inner horizon is replaced by a branch singularity.  Once again, \eqref{Q1} tells us this branch singularity is always to the left of the outermost horizon, and hence the black hole structure is preserved.

When the mass term in \eqref{BHPotKKToroidal} is non-decaying, $\da^2>2p/(p-1)$, the large $r$ asymptotes remain nevertheless non-singular. However, note that $\Lambda>0$, \eqref{l}. Firstly, when  $\tar<0$, the branch singularity occurs after the event horizon: the solution is always singular. When $0<2\tar\leq \ell^2$, the horizon is cosmological at finite $r_c$ and the metric is timelike beyond $r>r_c$. In either case we no longer have a black hole.

Note that the higher-derivative terms leave the near-horizon region of the near-extremal black hole unaffected: this is still AdS$_2\times \mathbf R_{(p-1)}$. This is not unexpected, given that our solution descends from and AdS black hole in higher dimensions.

%The Kaluza-Klein reduction may only be carried out for the specific values of $\da$ in equation \eqref{KKStatic2}: for positive integer values of $n$, $\da$ is restricted to $0\leq(p-1)\da^2/2<1$. Yet, inspecting the equations of motion derived from \eqref{CurvedKKGBAction} and the solution \eqref{MetriciKKToroidal}, nothing restricts the value of $\da$ \emph{a priori}, except the requirement that the solution should describe a black hole. This condition imposes that $\La<0$ and $(p-1)\da^2/2<p$, \cite{Charmousis:2009xr}. In particular, the range $1<(p-1)\da^2/2<p$ is perfectly fine and the limit $(p-1)\da^2/2=1$ (infinite $n$) is smooth, as in the Einstein case. In the former range, the black hole solution shares the properties of the planar AdS Gauss-Bonnet black hole it descends from. In the latter range, though no uplifting may be carried out, it was shown in \cite{Charmousis:2010zz} that the charged dilatonic Einstein solution shared the same thermodynamic properties as the asymptotically flat, spherical Reissner-Nordstr\"om black hole.

%Thus, by doing an analytic continuation on the parameter $\da$, we may obtain a substantially larger family of black hole solutions than simply dimensionally-reduced ones. Finally when we are at strong coupling $2\tar=l^2$ it is easy to note that we have generically an inner and outer event horizon with a branch singularity at $r>0$.

\subsection{Spherical dilatonic black hole\label{Section:CurvedEGBD}}

 We now move on to the scalar-tensor black hole with a curved horizon. As noted in section \ref{Section:GBBH}, in order for the higher-dimensional black hole to be a solution, the $n$-dimensional horizon must be a Gauss-Bonnet space, both its Ricci and Lanczos tensors must be proportional to the metric \eqref{EinsteinSpace}, \eqref{GBcond} respectively. The typical example is a maximally symmetric space with constant (positive, zero or negative) curvature, that is the sphere, the torus or the hyperbolic plane. A dimensional reduction of such a space would generically yield naked singularities at the poles of the reduced sphere  and along the radial direction, for the $(p+1)$-dimensional metric. A less trivial example of a Gauss-Bonnet space is that of an $(n+p-1)$-dimensional horizon, decomposed into a product of $m$-dimensional constant curvature spaces: $\l(\mathbf K^m\r)^{s+1}$, with $n+p-1=m(s+1)$ (see  Appendix \ref{Appendix:GBNonMaxHor} and \cite{Maeda:2010bu} for proof).

Starting from the seed $(n+p+1)$-dimensional Gauss-Bonnet black hole of section \ref{Section:GBBH} with potential \eqref{EGBBHPot} and horizon  sections $\l(\mathbf K^m\r)^{s+1}$, and identifying $\mathbf{\tilde K}^n\sim\left(\mathbf K^m\right)^s$ in the reduction Ansatz \eqref{KKStatic2}, we obtain the following dilatonic black hole solution with maximally symmetric sections $K^2_{(p-1)}$,
\bea
	\ud s^2_{(p+1)} &=&-f(r)\ud t^2+\frac{e^{\da\phi}\ud r^2}{f(r)}+r^2\ud K^2_{(p-1)},\label{MetricKKSph}\\
	f(r)&=&\bar \ka r^{(p-1)\da^2}+\frac{r^2}{\tar}\Bigg(1\mp\sqrt{1-2\tar\l(\frac{1}{\ell^2}+\frac{\tar \bar\Theta}{r^{4-2(p-1)\da^2}}-\frac{2\mf}{r^{p-\frac{p-1}2\da^2}}+\frac{\qf^2}{r^{2(p-1)}}\r)}\Bigg),\nn\\\label{BHPotKKSph}\\
	 e^{\phi}&=&r^{(p-1)\da}\,,\label{PhiKKSph}\\
	%h(r)&=&1+\frac{(p-2)(p-1)^2\da^2}{\l(1-\half(p-1)\da^2\r)}\ha k_{(p-1)}r^{-2+(p-1)\da^2}\,,\label{HarmFuncKKSph}\\
           A&=&-\sqrt{\frac{2(p-1)}{(p-2)+\frac{(p-1)}2\da^2}}\frac{ \qf}{r_+^{(p-2)+\frac{(p-1)}2\da^2}}\l[1-\l(\frac{r_+}{r}\r)^{(p-2)+\frac{(p-1)}2\da^2}\r]\ud t\,,\label{AKKCurved}
\eea
with the analytically continued coefficients \eqref{DilatonicReducedLovelockCoeff}, $\mf$ and $\qf$ the mass and charge integration constants and $\bar \ka$, $\bar \Theta$ related to the horizon geometry - to be defined shortly. In order to have a constant curvature horizon for the $(p+1)$-dimensional solution we have taken $m=p-1$, as well as traded the number of reduced dimensions $n=(p-1)s$ for the real parameter $\da$ as in \eqref{KKStatic2},
\be
	 s=\frac{(p-1)\da^2}{2-(p-1)\da^2}\,.
\ee
Once the Kaluza-Klein reduction has been carried out, we can analytically continue $\da$ to the whole real line, and enlarge the family of solutions obtained.

This black hole solution is a classical extremum of the action \eqref{CurvedKKGBAction} with the effective potential \eqref{effpot} for the scalar. Its coefficients are fixed by the reduction
\bea
	\tilde R&=&s(p-1)(p-2)\ka\,,\\
	\tilde\G&=&s(p-1)(p-2)\l[s(p-1)(p-2)-2(2p-5)\r]\ka^2\,,
\eea
and related to the curvature $\ka$ of the dilatonic black hole. We again concentrate on the Einstein branch of the solution. In the absence of $\tar$, the solution given here is in the Einstein frame \eqref{CurvedKKGBAction} and reduces to a solution discovered in \cite{Chan:1995fr}. In presence of the higher order corrections the term "Einstein frame" is ambiguous for the scalar and tensor parts mix inevitably through the higher order terms. For $\tar\neq0$, our solution \eqref{BHPotKKSph} is a generalisation of that of \cite{Chan:1995fr}.

Setting $\bar\ka =0$ in \eqref{MetricKKSph} brings us back to the case of a toroidal reduction and to the planar dilatonic black hole of section \ref{Section:PlanarEGBD}.
When $\bar\ka\neq0$, there are two horizon curvature terms in the solution \eqref{BHPotKKSph},
\bea
\bar\ka&=&\frac{(p-2)\ka}{\lambda \l(p-1-\la\r)}\,, \qquad \lambda=1-(p-1)\frac{\da^2}2\,,\qquad -\infty<\la<1\,, \label{tildek}\\
\bar\Theta &=& \frac{(p-1)(1-\la) \lambda\bar\ka^2}{\l(p-1-2\la\r)\l(p-1-3\la\r)(p-2)} \,, \label{ThetaKm}
\eea
where again $\kappa=0,\pm1$.
 Firstly, $\bar\ka$ stands for the standard curvature term, present for $\tar=0$, whereas $\bar\Theta$ originates from the higher order
Gauss-Bonnet condition  \eqref{GBcond}. For $p>3$, the sign of the curvature terms is dictated by $\lambda$, whereas for $p=3$ the last factor in the denominator of (\ref{ThetaKm}) also affects the sign. This will be important for the regularity of the solutions (\emph{cf} Tables \ref{TableM0La},\ref{TableM0La0}).
The $\bar\Theta$ dependence in \eqref{BHPotKKSph} is a blessing since the failure of the horizon curvature term $\bar\ka$ to be equal to $\kappa$ is translated in a solid deficit angle at $r=0$. In other words, the area of the $(p-1)$-dimensional horizon is reduced via the numerical factor in \eqref{tildek} with respect to that of the standard $(p-1)$-dimensional sphere (or hyperboloid). Hence, even for $(\mf,\qf)=(0,0)$, the solution presents a real curvature singularity at $r=0$ (including when $\tar=0$). Here, as we emphasized in section \ref{Section:EGBD}, the additional curvature term portrayed by $\bar\Theta$ can cover up the singularity by an event horizon without assuming matter in the face of $\mf$ (Tables \ref{TableM0La},\ref{TableM0La0}). This is a nice analytic example where highe-derivative curvature corrections in $\tar$ screen an otherwise naked singularity.

We now turn to the asymptotic region (large $r$).
For a start, we can have an idea of the solution in question by taking $\tar$ small (thus approaching the Einstein solution $\tar=0$) and expanding the square root of \eqref{BHPotKKSph}:
\be
f(r)\sim \bar \ka r^{(p-1)\da^2} + \frac{r^2}{\ell^2}+\frac{\tar \bar\Theta}{2 r^{2-2(p-1)\da^2}}-\frac{2\mf}{r^{p-2-\frac{p-1}2\da^2}}+\frac{\qf^2}{r^{2(p-2)}}+...
\ee
The solution asymptotes the $\tar=0$ limit while the leading $\tar$ term can create an additional horizon as we will see in a moment. This provides an overall effective picture of the solution, but let us go back to $\tar$ arbitrary.
Whenever $\lambda>0$, the asymptotes are driven as usual by the cosmological constant scale, $\ell$.  However, whenever $1-p<\lambda<0$, they are governed by the $\bar \ka$ and $\bar \Theta$ terms, a more exotic situation. This can be interpreted in the Einstein limit $\tar=0$ as the fact that the uplifted solution can be either an AdS black hole ($\la>0$) or an AdS$\times \mathbf H$ spacetime ($\la<0$), \cite{Gouteraux:2011qh}. Although this does not happen when the Gauss-Bonnet term is turned on, this interplay of asymptotes still occurs. The special value $\la=0$ is regular if one takes a scaling limit where simultaneously $\ka=0$, such that $\bar\ka$ remains finite but $\bar\Theta$ vanishes (all terms in the effective potential \eqref{effpot} collapse to a single exponential). Then, the planar solution is obtained. If $-\infty<\la<1-p$, the mass term dominates over the $\La$ term at infinity, but remains subdominant to the horizon curvature terms $\bar\ka$ and $\bar\Theta
 $, and the asymptotic nature of the spacetime is determined by the sign of $\bar\ka$. The limiting value $\la=1-p$ can be regularised by taking another scaling limit with $\La=0$ so that $\ell$ remains finite (the $\La$ term in the effective potential \eqref{effpot} drops out), and the asymptotes are still driven by the curvature terms.

%Proceeding along the lines of the previous sections \ref{Section:CurvedDiagKKRed} and \ref{Section:PlanarEGBD}, we reduce along an arbitrary number of the $\mathbf K^m$ composing the horizon, assuming that
%\be
%	n=sm\,, \quad p-1=m \Leftrightarrow \quad p+n-1=(q+s)m\,.
%\ee
%Projecting the Einstein's equation on the $(n+p-1)$-dimensional coordinates yields the relation
%\be
%	(p-2)\bar k = (n-1)\tilde k=(p+n-2)k
%\ee
%relating the normalised curvatures of, respectively, the $(p-1)$-dimensional horizon of the dilatonic black hole ($\bar k$), the $n$-dimensional constant curvature space $\mathbf K^n=\l(\mathbf K^m\r)^s$ which we have compactified  ($\tilde k$) and the $(n+p-1)$-dimensional original horizon ($k$).
%Moreover,
%\bea
%	 R&=&(n+p-1)(p+n-2)k=(s+1)(p-1)(p-2)\bk\,,\\
%	\G&=&(s+1)(p-1)(p-2)\l[(s+1)(p-1)(p-2)-(p-3)(p-4)\r]\bk^2\,.
%\eea
%Moreover,
%\bea
%	\Theta&=&\frac{\G}{(n+p-1)(n+p-2)(n+p-3)(n+p-4)}-\l(\frac{R}{(n+p-1)(n+p-2)}\r)^2\nn\\
%		&=&\frac{2(p-1)(p-2)s\bk^2}{\l[(p-1)(s+1)-1\r]^2\l[(p-1)(s+1)-2\r]\l[(p-1)(s+1)-3\r]}\,.
%\eea
%Even for a four-dimensional spacetime ($p=3$), the Gauss-Bonnet coupling $\tar$ engenders a non-trivial curvature contribution from the maximally symmetric horizon in the black hole potential through the $\Theta$ term.

To determine the presence of horizons and branch singularities, we can proceed as in section \ref{Section:PlanarEGBD} and look for roots of, respectively:
\bea
\Pp(x)&=&x^{2(p-1)}\left[\frac{2}{\ell^2}+2 \bar \ka x^{-2\lambda}+\tar (\bar \ka^2+2 \bar \Theta)x^{-4\lambda} \right]-4 \mf x^{p-1-\lambda}+2 \qf^2\,, \label{Pcurved} \\
\Q(x)&=&-\tar P(x)+x^{2(p-1)}(1+\tar \bar \ka x^{-2\lambda})^2\,. \label{Qcurved}
\eea
Already \eqref{Qcurved} tells us that $\Q(r_h)>0$ (for $\Pp>0$), which is particularily useful since we need to always verify $r_h>r_s$ in order for the solution to have a regular outer event horizon. As the polynomial $\Pp$ is obtained upon squaring $f(r)$, this imposes the additional inequality
\be
\tar \bar \ka r_h^{-2 \lambda}+1 \geq 0\,, \label{ConsistentHorizon}
\ee
which is redundant if $\tar \bar \ka\geq 0$, but otherwise bounds the permitted region of values for $r_h$.

\TABLE[t]{
{\footnotesize
\begin{tabular}{|c||c|c|c|c||c|c|c|c|}
	\hline
	&\multicolumn{4}{c||}{$\displaystyle \La<0 $} &\multicolumn{4}{|c|}{$\displaystyle\La>0$}\\ \hline
	&  \multicolumn{2}{c|}{$\displaystyle\tar>0$} &  \multicolumn{2}{|c||}{$\displaystyle\tar<0$} &  \multicolumn{2}{|c|}{$\displaystyle\tar>0$} &  \multicolumn{2}{|c|}{$\displaystyle\tar<0$}\\ \hline
	$\displaystyle \la$ &$\displaystyle\ka>0$ & $\displaystyle\ka<0$ & $\displaystyle\ka>0$ & $\displaystyle\ka<0$ & $\displaystyle\ka>0$ & $\displaystyle\ka<0$ & $\displaystyle\ka>0$ & $\displaystyle\ka<0$ 	\\ \hline
	$\displaystyle \left]-\infty,1-p\right[$ & $\bf N_c$ & $\bf O$ & $\bf C$ & $\bf O\; (N)$ & $\bf N_c$ & $\bf O$ & $\bf C$ & $\bf N$ \\ \hline
	$\displaystyle \left]1-p,0\right[$ & $\bf C$ & $\bf N$ & $\bf C$ & $\bf N$ & $\bf N_c$ & $\bf O$ & $\bf N_c$ & $\bf O\; (N)$ \\ \hline
	$\displaystyle 0$ ($\ka=0$) & \multicolumn{4}{c||}{$\bf N\; (\varnothing)$} &\multicolumn{4}{c|}{$\bf N_c\; (\varnothing)$} \\ \hline
	$\displaystyle \left]0,1\right[$ & $\bf N\; (\varnothing)$ & $\bf K,\, O\; (\varnothing)$ & $\bf O$ & $\bf O$ & $\bf C$ & $\bf C$ & $\bf O,\, C\; (\varnothing)$ & $\bf N_c$ \\ \hline
	$\displaystyle \left]2/3,1\right[$ (p=3) &  $\bf O\; (\varnothing)$ & $\bf O\; (\varnothing)$ & $\bf N$ & $\bf K,\,O$ & $\bf O,\, C$ & $\bf N_c$ & $\bf C\; (\varnothing)$ & $\bf C\; (\varnothing)$  \\ \hline
	$\displaystyle \da=0$ ($p>3$) & $\bf AdS\; (\varnothing)$ & $\bf K\; (\varnothing)$ & $\bf AdS$ & $\bf K$ & $\bf C$ & $\bf\varnothing$ & $\bf C\; (\varnothing)$ & $\bf \varnothing$  \\ \hline
\end{tabular}
}
\caption{Summary of horizon screening of the background ($\mf=0$, $\qf=0$) curvature singularity by the $\bar\Theta$ term, in terms of the sign of $\La$,  $\tar$ and $\bar\ka$ as well as the value of $\la$ ($\da^2$). The case $\la=1-p$ is displayed in Table \protect\ref{TableM0La0}. $\bf O$: Outer (event) horizon ; $\bf K$: Killing horizon ; $\bf C$: Cosmological horizon ; $\bf N_{(c)}$: (Cosmological) Naked singularity ; $\bf \varnothing$: undefined background. In parenthesis, we indicate that the situation depends on the value of $\tar\La$ given the dependence of \protect\eqref{BranchSing}, \protect\eqref{ThetaHorizon} on these factors. The last row gives the $\da=0$ case for comparison.}
\label{TableM0La}
}

Unlike the case $\bar \ka=0$, the horizon positions now depend on $\tar$ (the dependence on $\tar$ will be reiterated  by the black hole thermodynamics). Whenever $\lambda>0$ and given that $\Q(r_h)>0$, the branch singularity will be generically covered by an event horizon. For simplicity, we will discuss in detail the presence of horizons when the mass and charge parameters  $\mf$, $\qf$ are turned on in section \ref{section:GalBH} for the Galileon black hole. Here, we restrict to $\mf=0$, $\qf=0$, and determine to what extent the $\bar\Theta$ term can regularise the background curvature singularity by covering it with an event horizon.

\TABLE[t]{
\begin{tabular}{|c||c|c|c|c|}
	\hline
	$\displaystyle \La=0$&  \multicolumn{2}{c|}{$\displaystyle\tar>0$} &  \multicolumn{2}{|c|}{$\displaystyle\tar<0$} \\ \hline
	$\displaystyle \la$ &$\displaystyle\ka>0$ & $\displaystyle\ka<0$ & $\displaystyle\ka>0$ & $\displaystyle\ka<0$\\ \hline
	$\displaystyle \left]-\infty,0\right[$ & $\bf N_c$ & $\bf N_c$ & $\bf N_c$ & $\bf N$  \\ \hline
	$\displaystyle \left]0,1\right[$ & $\bf N$ & $\bf C$ & $\bf O$ & $\bf N_c$  \\ \hline
	$\displaystyle \left]2/3,1\right[$ (p=3) & $\bf O$  & $\bf N_c$ & $\bf N$ & $\bf C$   \\ \hline
	$\displaystyle \da=0$ ($p>3$) & $\bf Minkowski$  & $\bf \varnothing$ & $\bf Minkowski$ & $\bf \varnothing$   \\ \hline
\end{tabular}
\caption{Summary of horizon screening of the background ($\mf=0$, $\qf=0$) curvature singularity by the $\bar\Theta$ term for $\La=0$, in terms of the sign of  $\tar$ and $\bar\ka$ as well as the value of $\la$ ($\da^2$). The case $\da=0$ is displayed in Table \protect\ref{TableM0La}. Same conventions as in Table \protect\ref{TableM0La}.}
\label{TableM0La0}
}

First off,  for $(\mf,\qf)=(0,0)$, the basic condition $2\tar\leq \ell^2$, valid for $\bar\kappa=0$, is relaxed by the explicit $\bar\Theta$ dependence in the potential. The polynomials $\Pp$ and $\Q$ can even be solved exactly:
\bea
	r_s^{4\la}&=&\frac{2\tar\bar\Theta}{1-\frac{2\tar}{\ell^2}}\,, \label{BranchSing}\\
	r_h^{2\la}&=&\frac{-\bar\ka\ell^2}{2}\left[1\pm\sqrt{1-\frac{2\tar}{\ell^2}\left(1+\frac{2\bar\Theta}{\bar\ka^2}\right)}\right]\,.\label{ThetaHorizon}
\eea
These $r_s$ and $r_h$ are of course well-defined only for some ranges of $\da$ as well as some relative signs of $\tar$, $\bar\ka$, $\bar\Theta$ and $\Lambda$, and the inequality \eqref{ConsistentHorizon} imposes further constraints. We summarise the relevant cases in Tables \ref{TableM0La},\ref{TableM0La0}. Inspecting these results, one notes that contrarily to $\mf\neq0$, the background can be well-defined when $\la<1-p$, that is the curvature singularity is covered by a (geometrical) event horizon. This can also happen for $\bar\ka<0$ for all values of $\la$ and both signs of $\La\neq0$ and $\tar$ ; for $\bar \ka>0$ however, it only happens for $\la>0$ (with $\La=0$ included).

\subsection{Galileon black hole \label{section:GalBH}}

In this section, we will consider the Ansatz \eqref{KKGalileon} for the reduced Galileon action \eqref{CurvedGBGalileonAction}. For the sake of clarity and since we are interested for this section in four-dimensional GR modifications, we will restrict ourselves to $p=3$ space dimensions and switch off the electromagnetic field. The general Galileon solution for arbitrary $p$ and EM field is related to \eqref{BHPotKKSph} via a conformal transformation and a redefinition of the radial coordinate $R$, namely,
\be
e^{-\da \phi}=r^{-(p-1)\da^2}, \qquad R=\sqrt{n+1}r^\la
\ee
and can thus be reproduced with ease.

In $p=3$ space dimensions, the Galileon action \eqref{CurvedGBGalileonAction} admits the following classical black hole solution, which descends from the higher-dimensional Gauss-Bonnet black hole using the Ansatz \eqref{KKGalileon},
\bea
	\ud \bar s^2_{(4)} &=& -V(R)\ud t^2 + \frac{\ud R^2}{V(R)}+ \frac{R^2}{n+1} \ud \bar K^2_{(2)}\,, \label{GalileonMetric}\\
	V(R)&=& \ka+\frac{R^2}{\tar}\l[1\mp\sqrt{1-\frac{2\tar}{l^2}-\frac{2\tar^2 \ka^2}{(n-1)R^4}+\frac{4 \tar \mf}{R^{3+n}}}\r], \label{GalBHPot}\\
	\tar &=&  2\ha n (n+1), \qquad \frac{1}{\ell^2}=\frac{-2\Lambda}{(n+2)(n+3)} \qquad\\
	e^\phi&=&\frac{R^2}{n+1}\,, \label{GalPhi}
\eea
where $\ka=0,1,-1$ is the normalised horizon curvature and we have redefined for this section the constants $\tar$ and $\ell$. The parameter $n$ is analytically continued to the whole real line and obviously \eqref{GalileonMetric} has a higher-dimensional origin only for $n$ positive integer. Setting $\Lambda=0$ and therefore a spherical horizon $\ka=1$, we will start by making some qualititative remarks and describing properties of the solution without entering into technical details. Note that, taking carefully the $n=0$ limit switches off the scalar field and the higher-derivative corrections, and we obtain pure GR in \eqref{CurvedGBGalileonAction} and a Schwarzschild black hole \eqref{GalileonMetric}. This is particularly interesting since the scalar-tensor solution given above is a continuous deformation of the Schwarzschild solution. This will be true especially when $n$ is close to zero. Note also that the solid angle deficit disappears at $n=0$, otherwise the area of the horizon is given by $\frac{4\pi R^2}{n+1}$. As stressed in the previous sections this will give, at $R=0$, a true curvature singularity even if $\mf=0$. For large $R$, we have asymptotically a spacetime metric similar to that of a gravitational monopole, \cite{Barriola:1989hx}. Taking $\tar=0$, gives a standard Einstein dilaton solution with a Liouville potential. We expand \eqref{GalBHPot} for small $\tar$ and large $R$. We obtain,
\be
V(R)=1+\frac{\tar }{(n-1)R^2}-\frac{2\mf}{R^{n+1}}+...
\ee
This solution is reminiscent of a RN black hole solution where the role of the electric charge is undertaken by the horizon curvature correction in $\tar$. This term dominates the mass term close to the horizon and for $n<1$. Note that it can be of negative sign depending on the value of $n$ and $\tar$. As expected, the further we are from $n=0$, the further we deviate from a standard four-dimensional radial fall-off. This is the overall picture.

We  now analyse in some detail the horizon structure. As before, we define two polynomials:
\bea
\Pp(x)=-4 \mf+ \frac{\tar (n+1)}{n-1} x^{n-1}+2 x^{n+1}\,,\label{galhorizon}\\
\Q(x)=-\tar \Pp(x)+\l(x^{\frac{3+n}{2}}+\tar x^{\frac{n-1}{2}} \r)^2\,. \label{galbranch}
\eea
Horizons $R_h$ of \eqref{GalileonMetric} are roots of \eqref{galhorizon} whereas branch curvature singularities $R_s$ are roots of \eqref{galbranch}. Having squared the potential \eqref{GalBHPot} we need to also make sure that $R_h^2\geq \tar$.
The first important question we want to deal with is the central singularity at $R=0$, which is due to the solid deficit angle and is present even if $\mf=0$. Setting $\mf=0$, we have one branch singularity $R_s$ and a single event horizon $R_h$,
\be
\label{galvalue}
R_s^4=\frac{2\tar^2}{n-1},\qquad R_h^2=\frac{\tar(n+1)}{2(1-n)}\,.
\ee
Keeping in mind that $n=0$ is the GR limit (\emph{e.g.} Minkowski), we find that  for $-1<n<1$ and $\tar>0$ the singularity at $R=0$ is covered by an event horizon created by the higher-order curvature correction. In its absence ($\tar=0$), this solution would have been singular. The UV (small $R$) behaviour of the solution is regularised by the presence of the higher-order terms. If $n>1$ or $n<-1$, then $\tar<0$ is needed in order to preserve the event horizon. Here, when $n>1$ the curvature singularity is no longer at $R=0$ but at $R_s$ \eqref{galvalue} (with $R_s<R_h$ always).  The remaining cases are singular.

Now let us switch on the mass, $\mf \neq 0$. Whenever $\tar>0$, we have a single event horizon since $\Pp$ is increasing. When $-1<n<1$, there is no branch singularity however small $\mf$ is. On the contrary, when $n>1$, the mass is bounded from below in order to avoid a branch singularity:
\be
\mf> \l(\frac{2}{n+3} \r)^{\frac{n+3}{4}} \frac{\tar^{\frac{n+1}{2}}}{n-1}\,.
\ee
When $n<-1$, the solution is also a black hole but the mass term is not falling off at infinity. The region of most immediate interest is whenever $n$ is small but not zero.

The black hole properties are rather different for $\tar<0$. When $-1<n<1$, there is an inner and an outer event horizon as long as the following condition is fulfilled:
\be
\l(\frac{1}{2} \r)^{\frac{n+1}{2}}<\frac{\mf (1-n)}{|\tar|^{\frac{n+1}{2}}}<\l(\frac{2}{n+3} \r)^{\frac{n+3}{4}}.
\ee
When $n>1$, a single event horizon exists, covering a single branch singularity with $R_s<R_h$, \eqref{galbranch}.

\section{Holography for Einstein Gauss-Bonnet Dilaton theories \label{Section:HoloEGBD}}

\subsection{Holographic dictionary}

In this section, we describe how to set up the holographic dictionary for the set of Einstein Gauss-Bonnet Dilaton theories studied in section \ref{Section:CurvedDiagKKRed}. For simplicity, we restrict the discussion to toroidal reductions and flat boundaries. The derivation of the lower-dimensional dictionary \emph{via} generalised dimensional reduction is explained in full detail in \cite{Kanitscheider:2009as,Gouteraux:2011qh}, so we just sketch it here (see also the review \cite{Skenderis:2002wp}).

In the higher-dimensional, conformal theory, the most generic asymptotic form of the metric is given by the Fefferman-Graham expansion\footnote{One has to be careful with holographic renormalisation when higher derivatives are present, as they will generically source new relevant dual operators, \cite{Skenderis:2009nt}. This is evaded in the case of Gauss-Bonnet gravity since the field equations remain second order, but there might still be irrelevant modes propagating in the bulk if the higher derivatives are treated non-perturbatively. We will forget about such modes in the present analysis. We would like to thank Marika Taylor for pointing out this subtlety to us.}
\be \label{AdSFG}
	\ud s^2_{(n+p+1)} = \frac{\ud z^2}{4z^2} + \frac1zg_{ab}\ud z^a\ud z^b\,, \quad g = g_{(0)}+z^{(n+p)/2}\left(g_{(n+p)}+h_{(n+p)}\log z\right)+\ldots,
\ee
where indices $(a,b)$ run over boundary coordinates and we have suppressed these indices in the expression of $g$ (the boundary metric) for brevity. For a flat boundary, only the source, $g_{(0)}$, and the vev, $g_{(n+p)}$, appear at order lower than $(n+p)/2$ in the expansion. The log term $h_{(n+p)}$ is only present for even-dimensional boundaries, and is related to the metric variation of the holographic Weyl anomaly. For a curved boundary, all terms $g_{(i<(n+p)/2)}$ at intermediate orders appear, but can be determined from $g_{(0)}$ using the equations of motion.

The on-shell action contains UV divergences as $z\to0$, which can be removed by supplementing it with the appropriate counterterms. For a flat boundary, a single one is necessary, \cite{Brihaye:2008xu},
\be
	S_{ct} = \frac1{8\pi G_N}\int_{\partial\mathcal M}\ud^{n+p}x\sqrt{-\ga}\,\frac{(n+p-1)}{3L}\left(2+\sqrt{1-\frac{2\ta}{L^{2}}}\right),
\ee
where $\ga$ is the induced metric on the $(n+p)$-dimensional hypersurface $\partial M$. The renormalised action is then
\be
	S_{ren} = S + S_{GHY} + S_{M}+S_{ct}\,, \label{HoloGBAction}
\ee
where $S$ is given by \eqref{GBAction}, while $S_{GHY}$ and $S_M$ are defined on $\partial M$ to make the variational problem well-defined, \cite{Myers:1987yn}.

The holographic stress-energy tensor is the one-point function derived from the on-shell action \eqref{HoloGBAction} by functional derivation with respect to $g_{(0)}$:
\be
	\l<T_{ab}\r> = \frac2{\sqrt{-g_{(0)}}}\frac{\delta S_{ren}}{\delta g_{(0)}^{ab}}=\frac{(n+p)L_e^{n+p-1}}{16\pi G_N}g_{ab}^{(n+p)}+\ldots
\ee
where dots contain a possible contribution from the Weyl anomaly. Here, $L_e$ is the \emph{effective} radius of the background AdS spacetime, defined from \eqref{AdSEffLambda}. The CFT Ward identity is then just the statement that the expectation value of the stress-tensor is traceless, up to the conformal anomaly for even-dimensional boundaries:
\be
	\l<T_{a}^a\r> = \mathcal A\,.\label{WardIdentity}
\ee

Once a consistent dimensional reduction has been established, the lower-dimensional holographic dictionary is most easily derived in the dual frame, \cite{Boonstra:1998mp,Kanitscheider:2008kd,Kanitscheider:2009as,Gouteraux:2011qh}, where the AdS asymptotics are preserved in the lower-dimensional metric. Then, it has the same Fefferman-Graham expansion as \eqref{AdSFG}. In this frame, the action is \eqref{CurvedGBGalileonAction}. The expansion for the Kaluza-Klein scalar is fixed by the reduction; the counterterm action can be reduced; the lower-dimensional one-point correlation functions $\l<\bar T_{ij}\r>$ and $\l<\bar{\mathcal O}_\phi\r>$ follow from $\l<T_{ab}\r>$ projected along external or compact dimensions; the reduced Ward identity measures the departure from conformality, \cite{Kanitscheider:2009as}. In the frame \eqref{CurvedKKGBAction}, the counterterm action reads
\be
	\bar S_{ct} = \frac1{8\pi \bar G_N}\int_{\partial\bar{\mathcal M}}\ud^{p}x\sqrt{-\bar\ga}\,\frac{(p-1)}{3\ell}\left(2+\sqrt{1-\frac{2\tar}{\ell^{2}}}\right)e^{-\frac\da2\phi},\label{EGBDCounterterm}
\ee
with $\ell$ and $\tar$ given by \eqref{DilatonicReducedLovelockCoeff}. From the reduced holographic stress-energy tensor, one can for instance derive the thermodynamic and first-order hydrodynamic behaviour of a given family of black holes, which we proceed to do in sections \ref{Section:ThermoEGBD} and \ref{Section:HydroEGBD}. We stress that this procedure is valid for all solutions of the higher-dimensional theory respecting the symmetries of the Kaluza-Klein Ansatz \eqref{KKStatic2}, not just for the specific family of solutions in section \ref{Section:PlanarEGBD}.

\subsection{Thermodynamics\label{Section:ThermoEGBD}}

It is well-known that for a planar horizon, the thermodynamics of the Gauss-Bonnet black hole is identical to Schwarzschild-(A)dS: the Gauss-Bonnet coupling only couples to curvature terms. All the expressions for the thermodynamic quantities such as the temperature, entropy and free energy carry through from Einstein to Gauss-Bonnet, including of course the phase diagram. This is preserved by the reduction scheme. The thermodynamics of the dilatonic planar Einstein black hole was studied in detail in \cite{Charmousis:2010zz} in four dimensions (see \cite{Gouteraux:2011qh} for generic dimensions), and we refer to this work for a complete description of the canonical and grand-canonical ensemble. We shall simply recall their main characteristics in the grand-canonical ensemble here.

Given that the Kaluza-Klein reduction which yielded the dilatonic solution is diagonal, it is straightforward to derive all thermodynamic quantities in the lower-dimensional frame from the known higher-dimensional ones,  \cite{Myers:1988ze,Jacobson:1993xs,Cai:2001dz,Deser:2002jk,Cai:2003gr,Clunan:2004tb,Maeda:2010bu}. Note that the latter may be recovered by setting $\da=0$ in all subsequent formul\ae, obtaining the thermodynamics of the pure Gauss-Bonnet black hole in $p+1$ dimensions.

From the reduction Ansatz \eqref{KKStatic2} and \eqref{MetricKKSph}, the temperature can be deduced:
\bea
	 \mathfrak T &=&  \frac{r_+^{1-(p-1)\da^2/2}}{4\pi}\l[\frac{(p-(p-1)\da^2/2)}{\ell^{2}}-\left(p-2+(p-1)\frac{\da^2}2\right)\frac{\qf^2}{r_+^{2(p-1)}}\r].
\eea
Once one has determined the temperature, the other quantities can be deduced, since the on-shell action evaluated on the black hole solution is invariant, whether expressed in the higher or in the lower-dimensional frame:
\be
	\mathfrak S  = \frac{\V r_+^{p-1}}{4\bar G_N}
\ee
where we have defined $\V$ the volume of the horizon.
The charge and chemical potential are
\bea
	\mathfrak Q&=& \frac{\V\qf}{16\pi \bar G_N}\sqrt{2(p-1)\left(p-2+(p-1)\frac{\da^2}2\right)}\,,\\
	\Phi&=&\sqrt{\frac{2(p-1)}{\left(p-2+(p-1)\frac{\da^2}2\right)}}\frac{\qf}{r_+^{p-2+(p-1)\da^2/2}}\,,
\eea
consistent with \eqref{AKKCurved}. The mass is
\be
	\mathfrak M=\frac{\V(p-1)}{16\pi \bar G_N}\mf=\frac{\V(p-1)r_+^{p-(p-1)\da^2/2}}{16\pi \bar G_N}\l[\ell^{-2}+\frac{\qf^2}{r_+^{2(p-1)}}\r]
\ee
and finally the Gibbs free energy:
\bea
	\mathfrak{G}&=&\mathfrak M-\mathfrak T\mathfrak S -\Phi\mathfrak Q\\
	&=&\frac{\V r_+^{p-(p-1)\da^2/2}}{16\pi \bar G_N}\left(1-(p-1)\frac{\da^2}2\right)\l[-\ell^{-2}-\frac{\qf^2}{r_+^{2(p-1)}}\r].
\eea

Other, important quantities are the heat capacity at constant chemical potential, and the electric permittivity at constant temperature :
\be
	C_\Phi =\mathfrak T\l.\frac{\ud\mathfrak S}{\ud\mathfrak T}\r|_\Phi \geq 0\,,\qquad \epsilon_{\mathfrak T} = \l.\frac{\ud\mathfrak Q}{\ud \Phi}\r|_{\mathfrak T} \geq 0\,,
\ee
where the inequalities indicate the local stability of the black hole. We shall refrain from writing out their (lengthy, not particularly enlightening) expressions explicitly.

Finally, one can show that the first law holds as expected
\be
	\ud\mathfrak M =\mathfrak T\ud\mathfrak S + \Phi\ud\mathfrak Q
\ee
by checking that
\be
	\l.\frac{\partial\mathfrak G}{\partial\mathfrak T}\r|_\Phi = -\mathfrak S\,,\qquad \l.\frac{\partial\mathfrak G}{\partial \Phi}\r|_{\mathfrak T} = -\mathfrak Q\,.
\ee

The thermodynamic behaviour of the planar black holes can be separated into two ranges, \cite{Charmousis:2010zz}:
\begin{itemize}
	\item $0\leq \da^2\leq2/(p-1)$: A single branch exists, stable both locally and globally, for all values of the temperature and chemical potential.
	\item $2/(p-1)<\da^2\leq2p/(p-1)$: A single branch exists, unstable both globally and locally, for all values of the temperature and chemical potential.
\end{itemize}
This behaviour can be interpreted in the Kaluza-Klein picture : indeed, it was shown in \cite{Gouteraux:2011qh} that for $0\leq \da^2\leq2/(p-1)$, the charged $\ga=\da$ planar Einstein black holes could be derived from the charged, planar AdS black hole. Accordingly, the thermodynamics is preserved by the reduction, whether in the Einstein or Gauss-Bonnet case.

 As for the range $2/(p-1)<\da^2\leq2p/(p-1)$, the neutral dilatonic Einstein black holes descend from an (asymptotically flat) black $n$-brane wrapped on a torus. The charged version of these black branes in $p+n+1$ dimensions yields another family of Einstein dilatonic black holes (with $\ga\da=2/(p-1)$), which does not coincide with the $\ga=\da$ family (except for $\da^2=2/(p-1)$). Yet, it turns out that both families of charged solutions share the thermodynamics of the higher-dimensional black $n$-brane.

It is quite an interesting result that this similarity survives the introduction of a Gauss-Bonnet coupling $\ha\neq0$: at least in the Einstein, neutral case, the Kaluza-Klein interpretation relied on the fact that the action \eqref{CurvedKKGBAction} was invariant under \eqref{AdSFlatMap}. As is apparent from \eqref{CurvedKKGBAction}, this is nolonger true once $\ha\neq0$.

\subsection{First-order hydrodynamics\label{Section:HydroEGBD}}

It is also interesting to look at non-equilibrium transport coefficients. To do this, one may take advantage of the reduction formul\ae\ for the first-order hydrodynamics in \cite{Kanitscheider:2009as},\footnote{See \cite{Gouteraux:2011qh} for the case including a Kaluza-Klein vector.} and the actual value for the shear viscosity to entropy density for the charged Gauss-Bonnet black hole, \cite{Ge:2009eh}:\footnote{See \cite{Brigante:2007nu,Visc} for related work.}
\be
	\frac{\eta}{s}=\frac1{4\pi}\left[1-8\pi(n+p-3)\ha\frac{T}{\rho_+}\right] \label{GBShearViscosity}
\ee
which may induce violation of the so-called KSS bound, \cite{Kovtun:2004de}, for positive Gauss-Bonnet coupling. From the above and the results in \cite{Kanitscheider:2009as}, we may deduce that the lower-dimensional shear and bulk viscosities are:
\bea
	\frac{\bar\eta}{\bar s}&=&\frac1{4\pi}\left[1-8\pi\left(p-3+(p-1)\da^2\right)\ha \frac{\bar T}{r_+^{1-\frac{p-1}2\da^2}}\right]\label{ShearViscosity}\\
	\frac{\bar\zeta}{\bar\eta}&=&\da^2\,. \label{BulkViscosity}
\eea
The previous condition on the violation of the KSS bound, \cite{Kovtun:2003wp,Kovtun:2004de}, is preserved, which is expected as this is a dynamical constraint on the regularity of the higher-dimensional spacetime, \cite{Kanitscheider:2009as}. The ratio coincides with the result of \cite{Cai:2009zv} in $p=4$ and with the appropriate identifications as well as zero charge, even though the action considered by these authors is different (the higher-order derivative terms for the scalar are missing). This can be interpreted as the fact that the shear viscosity is captured by gravitational interaction terms (actually, the two-point correlation function of transverse gravitons), so that the higher-derivative scalar terms do not affect this quantity. Generically, one expects that the introduction of a coupling between a scalar relevant in the IR and higher-derivative operators should generate a temperature dependence of the ratio $\eta/s$,\footnote{We would like to thank U. G\"ursoy for discussion on this point.} \cite{Cai:2009zv}-\cite{Cremonini:2012ny}, such as is observed in the Quark-Gluon-Plasma for instance. The fact that the temperature dependence of \eqref{ShearViscosity} drops out once the charge is turned off can here of course be understood from the scale invariance of the higher-dimensional theory.

The second ratio gives the bulk viscosity, and saturates the bound proposed in \cite{Buchel:2007mf}:
\be
	\frac{\bar\zeta}{\bar\eta}=\frac2{p-1}-2\bar c_s^2\,,\label{BuchelBound}
\ee
where $c_s^2$ is the adiabatic\footnote{that is, at fixed ratio of entropy $s$ versus charge density $\rho$, \cite{landau:1987:fm}.} speed of sound defined from the pressure $P$ and energy density $e$ as
\be
	\bar c_s^2=\left.\frac{\partial \bar P}{\partial \bar e}\right|_{\frac{s}{\rho}} = \frac{1-\half(p-1)\da^2}{p-1}\,. \label{speedOfsound}
\ee
Note however that this is a kinematical constraint descending from the reduction, and that there is no guarantee it should hold. Indeed, the non-diagonal reduction considered in \cite{Gouteraux:2011qh} violates the bound \eqref{BuchelBound} (see also \cite{Buchel:2011uj}).

\section{Non-diagonal reduction of Einstein Gauss-Bonnet theories \label{Section:S1NonDiagKKRed}}

One may also perform a non-diagonal reduction of the usual Gauss-Bonnet action, thus generating a gauge field in the lower-dimensional action from the higher-dimensional metric:
\bea
	\ud s^2 &=&e^{2\al\phi}\ud \bar s^2+e^{-2(p-1)\al\phi}\l[\ud w-\mathcal A\r]^2\,, \quad \al=-1/\sqrt{2p(p-1)}\,,\label{KKTorNonDiagRed}\\
	\mathcal A&=& A_\mu\ud x^\mu\,.
\eea
The reduced action takes the following form, \cite{MullerHoissen1990709}:\footnote{The relation to the conventions of \cite{MullerHoissen1990709} is as follows: $n=p+2$, $\psi=e^{\phi}$, $\al\to-\al$. The definition for the $P$ tensor is given in the Appendix by \eqref{Ptensor}.}
\be \label{KKVectorAction}
\begin{split}
	\bar S&= \int \ud^{p+1}x\,\sqrt{\bar g}\l\{\bar R-\half\partial\phi^2-\frac14e^{\sqrt{\frac{2p}{p-1}}\phi}\F^2-2\La e^{-\sqrt{\frac2{p(p-1)}}\phi}\r.\\
	&+\ha e^{\sqrt{\frac{2}{p(p-1)}}\phi}\l[\bar\G_{(p+1)}+2\frac{(p-2)(p+1)}{p(p-1)}\bar G^{\mu\nu}\partial_\mu\phi\partial_\nu\phi+2(p-2)\Box\phi\partial\phi^2\r.\\
	&\l.+\half(p-2)(p-3)\l(\partial\phi^2\r)^2\r] + \frac{3\ha}{16}e^{(2p+1)\sqrt{\frac{2}{p(p-1)}}\phi} \l[\l(\F^2\r)^2 - 2\F^{\la}_{\phantom{1}\mu}\F^{\mu}_{\phantom{1}\nu}\F^{\nu}_{\phantom{1}\rho}\F^{\rho}_{\phantom{1}\la}\r]\\
	& +\ha e^{(p+1)\sqrt{\frac{2}{p(p-1)}}\phi}\l[-\half\bar P^{\la\mu}_{\phantom{2}\phantom{2}\phantom{1}\nu\rho}\F_{\la\mu}\F^{\nu\rho}-\frac{(p-2)(2p+1)}{p(p-1)}(\F_{\mu\nu}\partial^\nu\phi)^2\r.\\
	 &\l.\l.+\frac{(p-2)\F^{\la\mu}}{\sqrt{2p(p-1)}}\l(\partial^\nu\phi\nabla_\nu\F_{\la\mu}-2\partial_\mu\phi\nabla^\nu\F_{\la\nu}\r)+\frac{(p-2)(5p+1)}{4p(p-1)}\F^2\partial\phi^2\r]\r\}.
\end{split}
\ee
This reduction is not generalised, as it involves a single internal dimension. However, one could combine it with a diagonal one along the lines of \cite{Gouteraux:2011qh} to make it so. We shall not do so here as the $\mathbf S^1$ reduction will already be enough for our purposes.

%This procedure could of course be generalised on an $n$-dimensional torus, by reducing successively on each of the $n$ $\mathbf S^1$. Then, it would not be consistent to truncate to a single scalar sector, we would need to keep one for each reduced direction. Moreover, unless we turned off all Kaluza-Klein vectors but the last, we would generate an axion per vector at each step after the first, initial reduction. This goes far beyond the scope of this work, and we shall restrict ourselves to a single $\mathbf S^1$ direction.

%Another limitation of the above is that we have not proven formally the consistency of the reduction. Yet, one may note the following arguments in favour of consistency. First, the generic group theoretic argument by which the massive modes can be truncated is valid for an Abelian isometry group. Second, by reducing over an $\mathbf S^1$, the internal space trivially satisfies the Einstein and Gauss-Bonnet space conditions, since it has zero curvature. Finally, inconsistency often arises because of the choice of an overconstrained reduction Ansatz, incompatible with the equations of motion. Said otherwise, one has chosen a set of lower-dimensional fields incompatible with the symmetries of the reduction. In this case, all higher-dimensional metric elements are accounted for by the presence of the gauge and scalar field, or equivalently the symmetries of the reduction are exhausted by the symmetries attached to the gauge and scalar field (local $U(1)$ symmetry and rescalings, \cite{PopeLectureNotes}).

Following \cite{Gouteraux:2011qh}, let us boost (with parameter $\omega$) the planar GB black hole in $p+2$ dimensions\footnote{One could also imagine starting from an action already containing a Maxwell field, so that the Gauss-Bonnet black hole carries an electric charge. However, \emph{via} the boost, it would acquire a component along the $w$-direction, which upon reduction will give an axion field (that is, a scalar with a shift symmetry).}
\bea
	\ud s^2 &=&-\frac{f(\rho)}{\rho f_0h(\rho)}\ud \tau^2 + \frac{\ud \rho^2}{4\rho^2f(\rho)}+\frac1\rho\ud R^2_{(p-1)}+\frac{h(\rho)}\rho\l[\ud w-\mathcal A\r]^2\,, \label{BoostedEGBBB}\\
	\mathcal A&=& -\cotanh\w\l(1-\frac1{h(\rho)}\r)\ud\tau\,,\quad h(\rho)=1+\sinh^2\w\left(1-\frac{f(\rho)}{f_0}\right),\label{BoostedEGBMaxwell}\\
	f(\rho) &=&\frac{1}{\ta}\l[1-\sqrt{1-\frac{2\ta}{L^{\;2}}\l(1-\l(\frac{\rho}{\rho_+}\r)^{p+1}\r)}\r],
\eea
where we restrict the discussion to the Einstein branch and we have defined $f_0=f(\rho =0)$. Note that in the system of coordinates above, the boundary is at $\rho=0$ and the central curvature singularity at $\rho\to+\infty$. The lower-dimensional fields read:
\bea
	\ud \bar s^2&=&-\frac{f(\rho)\ud \tau^2}{f_0\rho^{\frac{p}{p-1}}h(\rho)^{\frac{p-2}{p-1}}}+\l[\frac{h(\rho)}\rho\r]^{\frac1{(p-1)}}\l[ \frac{\ud \rho^2}{4\rho^2f(\rho)}+\frac{\ud R^2_{(p-1)}}\rho\r],\label{BoostedEGBDBB}\\
	e^\phi &=&\l[\frac{h(\rho)}{\rho}\r]^{\sqrt{\frac{p}{2(p-1)}}}\,,\quad \mathcal A= -\cotanh\w\l(\frac{1}{\cosh^2\w}-\frac1{h(\rho)}\r)\ud\tau\,.\label{BoostedEGBDMaxwell}
\eea
This charged dilatonic Gauss-Bonnet black hole has an event horizon at $\rho_+$ and a curvature singularity at $\rho\to+\infty$. Another singularity, $\rho_c$, sits at the zero of $h(\rho)$
\be
	h(\rho_c)=0 \Longleftrightarrow \rho_c^{p+1}=\frac{\rho_+^{p+1}\sinh^2\al}{\frac{\ta}{2L^2}\frac{\cosh^4\al}{\sinh^2\al}-1} \quad \textrm{if} \quad  \frac{2\sinh^2\al}{\cosh^4\al}<\frac{\ta}{L^2}<1\,,
\ee
so only for a small enough boost parameter $\w$. This is a pure artifact of the Kaluza-Klein reduction, and it is resolved in the higher-dimensional picture (the boosted planar AdS black hole is of course regular everywhere, except at $\rho\to+\infty$), \cite{Gibbons:1994vm}.

In the Einstein limit $\ta\to0$, this reduces to a special instance of the family of planar black hole solutions studied in \cite{Charmousis:2009xr,Charmousis:2010zz,Gouteraux:2011ce,Gouteraux:2011qh}  ($\ga\da=1$, $\da=\pm1/\sqrt3$ in the conventions of these works). 

The neutral limit corresponds to $\w=0$, that is zero boost as expected.  On the other hand, one may take the scaling limit:
\be
	\w\to+\infty\,, \quad \rho\to  e^{\w}\rho\,,\quad \tau\to e^{2\w}\tau\,,\quad x^i\to e^\w x^i\,.
\ee
This zooms in on the near-horizon region of the near-extremal black hole (NHE limit). This limit is most explicit by changing coordinates to
\be
	\rho^{-(p+1)}=r^{p+1}-r_-^{p+1}\,,\quad   r_+^{p+1}=\rho_+^{-(p+1)}\cosh^2\w\,,\quad  r_-^{p+1}=\rho_+^{-(p+1)}\sinh^2\w\,,
\ee
which has the effect of disentangling the two special points $r_\pm$, and recovers the coordinate system used in \cite{Charmousis:2010zz} for the so-called $\ga\da=1$ solution in the Einstein limit. Then, the extremal limit is $r_+=r_-$, and its near-horizon region has the same scaling symmetries as the Einstein black hole. It displays hyperscaling violation, in the same sense as in \cite{Gouteraux:2011ce,Huijse:2011efDong:2012se}.

In \cite{Gouteraux:2011qh}, the lower-dimensional holographic dictionary was derived when the reduction generates a vector field. In particular, one may find there how to define the proper counterterms to obtain the lower-dimensional renormalised action, etc. Once this is done, the thermodynamics of the solution \eqref{BoostedEGBDBB} can be determined:
\bea
	\mathfrak T&=&\frac{(p+1)}{2\pi L^2\cosh\w\sqrt{f_0\rho_+}}\,,\quad \mathfrak S=\frac{\mathcal V\cosh\w}{4\bar G_N\rho_+^{\frac p2}}\,,\label{BoostedEGBDTemp}\\
	\mathfrak M&=&\frac{\mathcal V}{16\pi \bar G_NL^2\sqrt{f_0\rho_+^{p+1}}}\l(p\cosh^2\w~+1\r),\label{BoostedEGBDMass}\\
	  \Phi&=&\tanh\w\,,\quad \mathfrak Q=\frac{\mathcal V(p+1)\cosh\w\sinh\w}{16\pi \bar G_NL^2\sqrt{f_0}\rho_+^{p+1}}\,,  \label{BoostedEGBDCharge}\\
\mathfrak G&=&-\frac{\mathcal V}{16\pi \bar G_NL^2\sqrt{f_0\rho_+^{p+1}}}\,,\label{BoostedEGBDGibbs}
\eea
%\bar W&=&\frac{\tilde \omega\al_{0}}{32\pi\bar G_N \al_1\sqrt{f_0\rho_+^{p+1}}}\left[(p+1)\sinh^2\w-1\r]\label{BoostedEGBDHelmholtz},
which are respectively the temperature, entropy, mass, chemical potential, electric charge and Gibbs thermodynamic potentials of \eqref{BoostedEGBDBB}. Given our choice of normalisation for the boundary metric, these formul\ae\ are identical to their Einstein counterparts. All of them obey the usual first law of thermodynamics.
Note that the temperature goes to zero as the black hole shrinks, and also when the boost parameter $\w$ diverges, signalling the two extremal limits, neutral and charged.

In the grand-canonical ensemble, one can readily check that the chemical potential is a growing, positive function of the boost parameter $\w$, so that one can be traded with the other to do the thermodynamic analysis. Consequently, \eqref{BoostedEGBDTemp} defines the equation of state $\rho_+[\mathfrak T,\Phi]$ for the black hole radius, and it is straightforward to see that the grand-canonical ensemble admits a single branch of locally and globally stable black holes, as expected from the thermodynamic properties of the higher-dimensional AdS planar black hole.

To conclude this section, we turn to the hydrodynamics of the boundary field theory. The reduction formul\ae\ obtained in \cite{Gouteraux:2011qh} are still valid, as they are a kinematical result, which depends only on the reduction scheme and not on the specifics of the theory under investigation. Thus, the shear and bulk viscosities, heat and DC conductivities are given by:
\bea
    \bar \eta&=&\eta\cosh\w\,,\quad  \bar\zeta_s =\frac{2\bar\eta}{p}\left[\frac{1}{p-1}-\frac{\sinh^2\w\left((p-1)\cosh^2\w+p+1\right)}{\left((p-1)\cosh^2\w+1\right)^2}\right],  \label{ViscositiesBoost}\\
    \bar\kappa&=&\frac{\bar T\bar\eta}{\cosh^2\w}\,,\quad   \bar\sigma_{DC}=\frac{\bar\kappa}{\bar T}=\frac{\bar\eta}{\cosh^2\w} \,,\label{ConductivitiesBoost}
\eea
where $\eta$ is the neutral ($q=0$) limit of \eqref{GBShearViscosity}.
The KSS bound is of course still preserved by the reduction, but the bound on the ratio of bulk to shear viscosity is violated. This stems from the fact that we have turned on a non-normalisable mode for the lower-dimensional gauge field. Said otherwise, we have modified the asymptotic expansion of the higher-dimensional metric, and so the boundary symmetries are different from the static case. Since this bound is a kinematical statement, which is dependent on the reduction scheme, there is no reason it should still be saturated, let alone obeyed, once a KK vector is present.

Note that, as one might have expected from the dimensional reduction, the DC conductivity, which is the lower-dimensional two-point correlator for the zero mode of the boundary gauge field, is related to the lower-dimensional shear viscosity, which is itself the lower-dimensional two-point correlator for the zero mode of the boundary metric. This is consistent with their interpretation in the higher-dimensional picture: both are correlators for the zero mode of the higher-dimensional boundary metric, and so both stem from the higher-dimensional shear viscosity.

One can compare our result \eqref{ConductivitiesBoost} to that of \cite{Cai:2011uh}, where a higher-order correction to the Reissner-Nordstr\"om black hole $P_{\la\mu\nu\rho}\mathcal F^{\la\mu}\mathcal F^{\nu\rho}$ is considered for $p=3$. Our result agrees with theirs in the zero charge limit, after matching conventions and defining an effective coupling for the gauge field $g_F=2Le^{-\phi/\sqrt3}$ and an effective AdS radius $L_e=Le^{\phi/2\sqrt3}$. In particular, the dependence of the result of \cite{Cai:2011uh} on $\ha$ is correctly reproduced from the (neutral) five-dimensional $\eta/s$ ratio. For non-zero charge, the answers differ since our solution is not a correction to the Reissner-Nordstr\"om black hole.

\section{Discussion and Conclusions\label{Section:CCL}}

In this work we have shown how applying generalised dimensional reduction techniques, \cite{Kanitscheider:2009as,Gouteraux:2011ce,Gouteraux:2011qh}, allows to generate  a class of higher-derivative scalar(-vector)-tensor theories.  Its field equations are second order PDEs. The four-dimensional scalar-tensor version includes a significant part of the general Horndeski/Galileon theory \cite{Horndeski1974,Deffayet:2011gz}. A non-diagonal reduction leads to an even more complex scalar-tensor vector theory that generalizes Einstein Maxwell (although here the reduction is rather restrictive). Furthermore, in arbitrary dimensions we have a holographic extension of Einstein Maxwell Dilaton theories of the type studied recently in \cite{Charmousis:2010zz} that describes the low energy behaviour at finite (rather than infinite) 't Hooft coupling. The theory we have studied has the following characteristics: it is parametrised by a continuous real parameter; it comprises higher-derivative couplings (Galileons) between the metric and scalar field; it admits up to three exponential potentials which have a higher-dimensional geometric origin. The reduction has the remarkable feature that consistency imposes both the Einstein and Gauss-Bonnet space conditions on the internal space, while other cases do not seem to lead to equations of motion deriving from a lower-dimensional action.

We have then presented a class of \emph{exact} (scalar-vector-tensor) lower-dimensional solutions, stemming from known higher-dimensional charged Einstein Gauss-Bonnet black holes. The higher-dimensional solutions can have the usual maximally symmetric sections \cite{Boulware:1985wk,Wheeler:1985qd,Wiltshire:1985us,Cai:2001dz}. This is in particular helpful in order to obtain planar horizon scalar-tensor black holes.
Indeed we find that for a planar horizon, the higher-derivative solution is quite similar to the Einstein Dilaton case. The horizon positions are independent of the Gauss-Bonnet coupling and hence so is the temperature, both in the higher and lower-dimensional theory.

For a scalar-tensor black hole with a spherical horizon  there are important changes. In fact, for the higher-dimensional black holes we need to consider non-homogeneous horizons, \emph{i.e.} products of spheres, so  that the lower-dimensional KK version has spherical (or hyperbolic) sections. These have to be treated with special caution for they are generically singular at $r=0$ in absence of a mass term.  This is due to the fact that the horizon sphere has a solid angular deficit, similar to the gravitational monopole solution, \cite{Barriola:1989hx}.  For zero Gauss-Bonnet coupling, this curvature singularity only disappears for zero deficit angle, which recovers the Schwarzschild solution. Turning on the Gauss-Bonnet coupling plays an essential role, unlike for planar horizons. On top of changing the horizon positions, it also creates additional inner or event horizons. Our result is that the first order curvature correction, in the guise of the Gauss-Bonnet coupling, quite generically dresses the naked singularity by a regular event horizon, \emph{even for zero mass}.  This is a non-trivial and rare example where higher-derivative corrections regularize a singular situation of lower order Einstein (Maxwell Dilaton) theory (though see also \cite{Dabholkar:2004dq} in a different setting).  The higher-derrivative correction acts as a stringy dressing to shield the singularity. We have reviewed the cases in which this happens in the lower-dimensional setup in Tables \ref{TableM0La} and \ref{TableM0La0}. Finally, we have analysed a black hole solution in the Galileon frame, which is conformally related to the previous frame. This is the first exact Galileon black hole which is also continuously related, via a real action parameter, to the GR black hole. Again, higher-derivative corrections induce extra horizons and the solutions are more regular than the lowest order scalar-tensor solutions. No Vainshtein effect is detected as one expects in the absence of matter.{\footnote{We thank E. Babichev, C. Deffayet and G. Esposito-Farese for pointing this out}}

Focussing on toroidal reductions, we have sketched how to set up the holographic dictionary for the lower-dimensional theory, and derived the thermodynamics and first-order transport coefficients for the planar dilatonic black holes. As could have been anticipated, they display the same thermodynamics as their Einstein counterparts. Both shear and bulk viscosities can be worked out using the reduction formul\ae. In spite of the presence of a relevant scalar field, they do not feature any temperature dependence at zero charge density. This is expected given their link to a higher-dimensional scale invariant theory.

When one reduces non-diagonally over a circle instead of diagonally over a compact space, higher-derivative couplings involving the lower-dimensional gauge field are produced; however, the reduction is not generalised, and has no continuous parameter (in \cite{Gouteraux:2011qh}, an example of generalised reduction is presented when one combines both a non-diagonal circle reduction and a diagonal generalised toroidal reduction). An exact black hole solution of the equations of motion can be exhibited, stemming from a planar Gauss-Bonnet black hole carrying a wave. Once more, the reduction allows to derive the holographic dictionary and transport properties.

Perspectives for future work are numerous, and include very different directions. On the gravity side, the higher-dimensional picture allows to gain intuition in order to try to find more solutions of the lower-dimensional setup. A simple extension should include a study of $p$-forms in Gauss-Bonnet gravity, \cite{Bardoux:2010sq}, and then of their dimensional reduction. This was performed in \cite{Gouteraux:2011ce} in the Einstein case. Another, rather involved but very interesting extension would be to include higher order Lovelock terms and then obtain the lower-dimensional scalar-tensor action. Some black hole solutions for Lovelock theory are known \cite{Myers:1988ze} and extensions to \cite{Bogdanos:2009pc} could be pursued with not too much difficulty. The Kaluza-Klein reduction would then give higher order terms such as the Paul term discussed in the Fab 4 self-tuning scenario. For toroidal black holes it would be interesting to know if it is possible to have Ricci flat horizons which are non homogeneous. In this way one could obtain toroidal black holes with differing thermodynamics to their lower order solutions. Last but not least, symmetries of the action appearing at zero higher order coupling $\ha$, if restored, could yield black $p$-brane solutions in Lovelock theory. 

In section \ref{Section:HoloEGBD}, we have restricted our study to the case with flat boundary. The case with curved boundary is somewhat more involved: the number of holographic counterterms grows with the dimension of spacetime, making the evaluation of the on-shell renormalised action computationally more complicated. Though this can (and has) been done in a number of cases, it deserves a closer study, as some issues arise, specific to Gauss-Bonnet theories. The presence of a geometrical horizon in the background of the Gauss-Bonnet black holes with a Gauss-Bonnet space horizon geometry makes it unclear which background should be used in thermodynamic studies: the $\mf=\qf=0$ spacetime, which has non-zero temperature and thus induces a conical singularity in the Euclidean action, or the extremal black hole $\mf_e\neq0$ with zero temperature? A similar issue  for hyperbolic Einstein black holes was resolved using the holographically renormalised action (see for instance \cite{Emparan:1999pm}), and it would be interesting to carry out the same analysis in our case. Moreover, this might also shed some light on the negative entropy problem, \cite{Clunan:2004tb}.

From the point of view of gauge-gravity duality, Einstein Maxwell Dilaton theories have proven to be interesting bottom-up setups either in the modeling of Condensed Matter systems or of QCD-like gauge theories, \cite{hQCD}. In this context, Gauss-Bonnet corrections allow to depart from strong coupling, and getting an analytic handle on these is of great importance. In holographic studies of Condensed Matter Systems, a recurring question is whether the AdS$_2\times\mathbf R^2$ near-extremal geometry of the Reissner-Nordstr\"om black hole is the true ground state of the theory or not. The fact that it has finite entropy at extremality argues against this. This issue can be resolved in the generic Einstein Maxwell Dilaton setup, \cite{Charmousis:2010zz,kachru,EMDHolo}. Then, the low energy fixed point displays hyperscaling violation, \cite{Gouteraux:2011ce,Huijse:2011efDong:2012se}. Both reductions of section \ref{Section:CurvedDiagKKRed} and \ref{Section:S1NonDiagKKRed} preserve the scaling symmetries of the Einstein fixed point. As in the Einstein case, it is expected that a much broader range of low energy scalings is possible once other higher-dimensional cases are considered. More generically, one needs to decide on the set of higher-derivative operators to include for the scalar and the gauge field, as these will drive the dynamics to different fixed points, as examplified above. It would be very interesting to study to what extent the generic hyperscaling symmetries of the low energy fixed points of the Einstein Maxwell Dilaton setup survive the inclusion of higher-derivative operators.

\acknowledgments

We gratefully acknowledge multiple discussions with E. Babichev, M. Caldarelli, U. G\"ursoy, F. Nitti, M. Paulos, K. Skenderis and M. Taylor. B.G. also wishes to thank K. Van Acoleyen and J. Van Doorsselaere for clarifications on their results. B.G. thanks the Laboratoire de Math\'ematiques et Physique Th\'eorique at Univ. Fran\c{c}ois Rabelais, Tours, France. C.C. and B.G. thank the Crete Center for Theoretical Physics at Univ. of Crete, Heraklion, Greece, for hospitality over the course of this work.
This work was partially supported by the European grants FP7-REGPOT-2008-1: CreteHEPCosmo-228644,
PERG07-GA-2010-268246 and the EU program "Thales" ESF/NSRF 2007-2013. This work was partially supported by the ANR grant STR-COSMO ANR-09-BLAN-0157. It was also cofinanced by the European Union (European Social Fund, ESF) and Greek national funds
through the Operational Program Education and Lifelong Learning" of the National Strategic Reference
Framework (NSRF) under Funding of proposals that have received a positive evaluation in the 3rd and
4th Call of ERC Grant Schemes

\appendix

\section{Diagonal Kaluza-Klein reduction \label{App:DiagonalKKRed}}

In this work, we shall only consider reductions for which the isometry group the the internal space is Abelian. These will be either diagonal toroidal reductions, reductions over Einstein spaces (typically, spheres) and non-diagonal $\mathbf S^1$ reductions. When the isometry group is Abelian, the massive modes of the Kaluza-Klein reduction can be truncated under the following group theoretic argument, \cite{PopeLectureNotes}: the zero modes transform as singlets under the action of the group, while the massive modes transform as doublets; if the group is Abelian, these cannot mix, and thus the massive modes are decoupled from the zero modes and can be truncated away. For curved internal spaces (spheres), the situation is quite more complicated if one insists on keeping (some of) the Kaluza-Klein vectors. Then, there are very few cases where the massive modes can be truncated, and they reduce to the sphere reductions of supergravity theories (see for instance \cite{Cvetic:1999xp}). In the cases we consider, the massive modes are not sourced and can be truncated.

We consider a Kaluza-Klein diagonal reduction of the action \eqref{GBAction}, with metric Ansatz:
\bea
	\ud s^2_{(p+n+1)}&=&\e^{2\al\phi}\ud \bar s^2_{(p+1)} + \e^{2\ba\phi}\ud \tilde K^2_{(n)}
	\label{KKStatic1}\,,\\
	A_M&=&(A_\mu(x^\mu),A_i=0)
	\label{AnsatzVector}\,,
\eea
where $\ud \tilde K^2_{(n)}$ is the metric of a generic $n$-dimensional Euclidean space. From now on in this section, we shall drop the subscripts indicating the dimensionality in parenthesis: bare quantities are $(n+p+1)$-dimensional, barred ones  $(p+1)$-dimensional while tilded ones $n$-dimensional.

Then, one finds that
\bea \label{ReducedEinsteingeneric}
	\sqrt{-g}R &=&\sqrt{-\bar g}\e^{((p-1)\al+n\ba)\phi}\l[ \bar R+\e^{2(\al-\ba)\phi}\tilde R-2(p\al+n\ba)\Box\phi\r.\\
	&&\l.-\left(p(p-1)\al^2+n(n+1)\ba^2+2n(p-1)\al\ba\right)\partial\phi^2\r].\nn
\eea
Setting the overall conformal factor in $\exp[\#\phi]$ in the action to unity requires
\be \label{EinsteinFrame}
	n\ba=(1-p)\al
\ee
upon which
\be \label{ReducedEinsteingeneric2}	\sqrt{-g}R =\sqrt{-\bar g} \left[\bar R-2\al\Box\phi-(p-1)\frac{p+n-1}{n}\al^2\partial\phi^2+\e^{2\frac{p+n-1}{n}\al\phi}\tilde R\r]\,.
\ee
After throwing away some total derivatives, the reduction of the Gauss-Bonnet term gives:
\be \label{ReducedGBgeneric}
\begin{split}
	\G=&e^{-4\al\phi}\l\{\bar\G-4\bar G^{\mu\nu}\partial_\mu\phi\partial_\nu\phi\l[(p-2)(p-3)\al^2+2n(p-2)\al\ba+n(n-1)\ba^2\r]\r.\\
		&\qquad\qquad\qquad-2\partial\phi^2\square\phi\l[(p-1)(p-2)(p-3)\al^3+3n(p-1)(p-2)\al^2\ba\r.\\
		&\qquad\qquad\qquad\qquad\qquad\qquad+\l.3n(n-1)(p-1)\al\ba^2+n(n-1)(n-2)\ba^3\r]\\
		&\qquad\qquad\qquad-\l(\partial\phi^2\r)^2\l[(p-1)(p-2)^2(p-3)\al^4+4n(p-1)(p-2)^2\al^3\ba\r.\\
		&\qquad\qquad\qquad\qquad\qquad\qquad\qquad\qquad+2n(p-1)(3np-2p-5n+3)\al^2\ba^2\\
		&\qquad\qquad\qquad\qquad\qquad\l.\l.+4n(n-1)^2(p-1)\al\ba^3+n(n-1)^2(n-2)\ba^4\r]\r\}\\
		&+2e^{-2(\al+\ba)\phi}\tilde R\l\{\bar R+\l(\partial\phi\r)^2\l[p(p-1)\al^2+2p(n-2)\al\ba+(n-3)(n-2)\ba^2\r]\r\}\\
		&+\tilde\G e^{-4\ba\phi}.
\end{split}
\ee
In order to have a canonically normalised kinetic term for the scalar (in the Einstein limit), we then set
\be \label{CanonicalScalar}
	\al=-\sqrt{\frac{n}{2(p-1)(n+p-1)}}=-\frac\da2\quad \Leftrightarrow \quad \da=\sqrt{\frac{2n}{(p-1)(p+n-1)}}\,.
\ee

For conciseness, we shall refrain from giving the full formul\ae\ of the reduced Lanczos and Einstein tensors in terms of $\al$, $\ba$ and $n$, and the reduced equations of motion are presented below in terms of $\da$ and after imposing \eqref{EinsteinFrame} and \eqref{CanonicalScalar}. The formul\ae\ \eqref{ReducedGBgeneric} and \eqref{ReducedEinsteingeneric} are useful however to compare with other results in the literature.

After imposing \eqref{EinsteinFrame} and \eqref{CanonicalScalar}, the metric Ansatz becomes:
\be
	\ud s^2 = e^{-\da\phi}\ud \bar s^2 +  e^{\frac{\phi}{\da}\l(\frac2{p-1}-\da^2\r)}\ud \tilde K^2\,,\qquad \frac{p-1}2\da^2 = \frac{n}{n+p-1}\,. \label{KKStatic3}
\ee
Exchanging the number of reduced dimensions for the parameter $\da$ and analytically continuing the latter to the whole real line is what makes this reduction generalised, \cite{Kanitscheider:2009as,Gouteraux:2011qh}.

We  first examine the projection of the equations of motion \eqref{GBEOM} on the external coordinates.
For this, it is useful to define what in four dimensions corresponds to the double dual tensor, \cite{Lanczos:1962zz}. Namely,
\be \label{Ptensor}
	P_{ABCD} = R_{ABCD}+R_{BC}g_{AD}-R_{BD}g_{AC}-R_{AC}g_{BD}+R_{AD}g_{BC}+ Rg_{(A|C|}g_{B)D}\,,
\ee
in terms of which the Lanczos tensor greatly simplifies
\be
	H_{AB}=R_{ACDE}P_B^{CDE}-\half g_{AB}\G\,.
\ee
The tensor (\ref{LanczosTensor}) is identically zero in four dimensions and the above equation is then a Lovelock identity. The $P$ tensor has the same symmetries as the Riemann tensor, is divergence-free and its trace is the Einstein tensor $P^A_{BAC}=G_{BC}$.  In four dimensions we have,  $P^{AB}_{\phantom{2}\phantom{2}\phantom{2}CD}=\star \bar R\star^{AB}_{\phantom{2}\phantom{2}\phantom{2}CD}=-\frac14\eps^{ABKL}\eps_{CDMN}R^{MN}_{\phantom{2}\phantom{2}\phantom{2}KL}$, the double dual to Riemann curvature. With the help of this definition, we are led to the reduced lower-dimensional Einstein equations for the metric:
\bea
	0&=&\bar{\mathcal E}^{\nu}_{(1)\mu}+\ha e^{\da\phi}\bar{\mathcal E}^{\nu}_{(2)\mu}\,, \label{ReducedEinsteinEq}\\
	\bar{\mathcal E}^{\nu}_{(1)\mu}&=&\bar G_{\mu}^{\nu}-\frac{\partial_\mu\phi\partial^\nu\phi}2-\frac{e^{\da\phi}}2F_{\mu\rho}F^{\nu\rho}+\left(\La e^{-\da\phi}+\frac{\partial\phi^2}4-\frac{e^{-\frac{2\phi}{(p-1)\da}}}2\tilde R+\frac{F^2}8e^{\da\phi}\right)\da_{\mu}^{\nu}\,,\\
 \bar{\mathcal E}^{\nu}_{(2)\mu}&=&-\frac{\tilde\G}2e^{\frac{-4\phi}{(p-1)\da}}\da_\mu^\nu+2\tilde Re^{\frac{-2\phi}{(p-1)\da}}\left[\bar G_{\mu}^\nu+\frac{(p-1)\da^2-2}{(p-1)\da}\left(\Box\phi\da_\mu^\nu-\nabla_\mu\nabla^\nu\phi\right)\right.\\
		 &&\left.+\left(\frac2{(d-1)^2\da^2}-\frac{1+\da^2}2\right)\partial_\mu\phi\partial^\nu\phi+\right(3\da^2+\frac{d-9}{d-1}+\frac{4}{(d-1)^2\da^2}\left)\frac{\partial\phi^2}4\da_\mu^\nu\right]\nn\\
		&&-\bar H_\mu^\nu+(1-\da^2)\bar G_\mu^\nu\partial\phi^2+2(1+\da^2)\bar P_{\mu\rho}^{\phantom{1}\phantom{1}\nu\sigma}\partial^\rho\phi\partial_\sigma\phi+2\da\bar P_{\mu\rho}^{\phantom{1}\phantom{1}\nu\sigma}\nabla^\rho\nabla_\sigma\phi\nn\\
	 &&+(1-\da^2)\left[\left(\Box\phi^2-(\nabla_\rho\nabla_\sigma\phi)^2\right)\da_\mu^\nu-
2\left(\Box\phi\nabla_\mu\nabla^\nu\phi-\nabla_\mu\nabla_\rho\phi\nabla^\nu\nabla^\rho\phi\right)\right]\nn\\
&& +\da\left(\da^2+1-\frac2{(d-1)\da^2}\right)\left(\nabla_\rho\nabla_\sigma\phi\partial^\sigma\phi\da_\mu^\nu-
\nabla_\mu\nabla_\rho\phi\partial^\nu\phi-\nabla^\nu\nabla_\rho\phi\partial_\mu\phi\right)\partial^\rho\phi\nn\\
&&+\da(1-\da^2)\partial\phi^2\left(\Box\phi\da_\mu^\nu-\nabla_\mu\nabla^\nu\phi\right)+
\left(\da(1+\da^2)-\frac2{(p-1)\da}\right)\Box\phi\partial_\mu\partial^\nu\phi\nn\\
&&+\frac{\partial\phi^2}2\left[\left(5\da^2+\frac{d-3}{d-1}-\frac{4\da^{-2}}{(d-1)^2}\right)\frac{\partial\phi^2}4\da_\mu^\nu-\left(\da^2+\frac{d+1}{d-1}-\frac{4\da^{-2}}{(d-1)^2}\right)\partial_\mu\phi\partial^\nu\phi\right].\nn
\eea

Turning now to the projection of \eqref{GBEOM} on the internal coordinates, we have that schematically:
\be\label{ReducedEinsteinEqInternal}
\begin{split}
	0=&\,2\ha e^{\frac{-\phi}{(p-1)\da}(4-(p-1)\da^2)}\tilde H^a_b +e^{\frac{-2\phi}{(p-1)\da}}\tilde G^a_b\left(-2+\ha\bar K_1e^{\da\phi}\right)\\
	&+\tilde\da^a_b\left[\bar R-\frac12\partial\phi^2+\frac{2\Box\phi}{(p-1)\da}-2\La e^{-\da\phi}-\frac18F^2e^{\da\phi}+\ha\bar K_2 e^{\da\phi}\right]. 
\end{split}
\ee
Taking the trace of \eqref{ReducedEinsteinEqInternal}, multiplying by $\da_b^a$ and subtracting back from \eqref{ReducedEinsteinEqInternal}, we obtain
\be
	0=\ha e^{\frac{-\phi}{(p-1)\da}(4-(p-1)\da^2)}\left(\tilde H^a_b-\frac{n-4}{2n}\tilde\G\tilde\da^a_b\right)+e^{\frac{-2\phi}{(p-1)\da}} \left(\tilde G^a_b+\frac{n-2}{2n}\tilde R\tilde\da^a_b\right)\left(\frac\ha2\bar K_1e^{\da\phi}-1\right),
\ee
where $\bar K_1$ and $\bar K_2$ are complicated expressions which depend solely on barred quantities and on external coordinates. Since the equation above is made of terms where the factors depend on separate variables, a separation of cases ensues:
\begin{itemize}
	\item Case 1:
\bea
	\tilde G_{a}^{\phantom{1}b}&=&-\frac{n-2}{2n}\tilde R\da_a^b\,,\label{EinsteinSpace1}\\
	\tilde H_{a}^{\phantom{1}b}&=&\frac{n-4}{2n}\tilde \G\da_a^b\,,\label{GBSpace}
\eea
generically constrains the internal space to be both an Einstein space and a Gauss-Bonnet space.
Flat Euclidean spaces obviously belong to this class, but it also allows for curved spaces such as the product of identical spheres, more interestingly. This case allows a consistent Kaluza-Klein reduction, as we shall see shortly. This is precisely this feature we shall exploit, in order to derive analytical exact black hole solutions of the reduced theory.
	\item Case 2:
\bea
	\ha \bar K_1&=&2e^{-\da\phi}\,,\label{Case2Constraint}\\
	\tilde H_{a}^{\phantom{1}b}&=&\frac{n-4}{2n}\tilde \G\da_a^b\,.
\eea
In that case, one needs to find an internal space which is Gauss-Bonnet without being Einstein (otherwise we fall back to Case 1). It is unclear whether such spaces exist, as the few known examples of Gauss-Bonnet spaces are also Einstein. Assuming such a space can be found, the system of equations of motion is overconstrained; three independent scalar equations need to be obeyed: one stems from the trace of \eqref{ReducedEinsteinEq}, another from \eqref{Case2Constraint} and a third one from the trace of \eqref{ReducedEinsteinEqInternal}. This is incompatible with the derivation of these equations of motion from an action principle involving a metric and a scalar field only.
	\item Case 3:
There exists a real number $\la$ such that
\bea
	\tilde H^a_b-\frac{n-4}{2n}\tilde\G\tilde\da^a_b &=& \la \left(\tilde G^a_b+\frac{n-2}{2n}\tilde R\tilde\da^a_b\right)\\
	\ha \bar K_1&=&2e^{-\da\phi}-\ha\la e^{\frac{-2\phi}{(p-1)\da}}\,.
\eea
\end{itemize}

This kind of casuistics is familiar in studies of the integrability of the Gauss-Bonnet equations of motion, in presence of sources or not, \cite{Charmousis:2002rc,Zegers:2005vx,Bogdanos:2009pc,Bardoux:2010sq}.

In the remainder, we shall focus on Case 1, and show that the resulting equations of motion derive from the reduction of the action. Before proceeding on, let us note that it is quite remarkable that the consistency of reduction of the Einstein Gauss-Bonnet equations of motion imposes an extra condition on the internal space. In contrast, as noted in the appendix of \cite{Gouteraux:2011qh}, the reduction from Einstein theory imposes the Einstein space condition on the internal space.

To find the equation of motion for the scalar, one now needs to take the trace of \eqref{ReducedEinsteinEq} and combine it linearly with \eqref{ReducedEinsteinEqInternal}. After some manipulations, one obtains:
\bea
	0&=&\tilde{\mathcal E}^{\phi}_{(1)}+\ha e^{\da\phi}\tilde{\mathcal E}^{\phi}_{(2)}\,, \label{ReducedPhiEq}\\
	\tilde{\mathcal E}^{\phi}_{(1)}&=&\Box\phi+2\da\La e^{-\da\phi}-\frac{2\tilde R}{(p-1)\da}e^{-\frac{2\phi}{(p-1)\da}}-\frac\da4F^2e^{\da\phi},\\
	\tilde{\mathcal E}^{\phi}_{(2)}&=&-\frac{4-(p-1)\da^2}{(p-1)\da}\tilde \G e^{\frac{-4\phi}{(p-1)\da}}+\tilde Re^{\frac{-2\phi}{(p-1)\da}}\left[2c_5\Box\phi-\frac{2-(p-1)\da^2}{(p-1)\da}\left(c_5\partial\phi^2+2\bar R\right)\right]\nn\\
			&&+\da\bar\G +2\bar G_{\mu\nu}\left[(c_4-\da c_2)\partial^\mu\partial^\nu\phi-c_2\nabla^\mu\nabla^\nu\phi\right]+\partial\phi^2\left[4\Box\phi+\da(\da c_4-3c_3)\partial\phi^2\right]\nn\\
			&&+c_4\bar R\partial\phi^2+2\left(\nabla^\mu\nabla^\nu\phi\nabla_\mu\nabla_\nu\phi-\Box\phi^2\right)+4(\da c_4-2c_3)\nabla_\mu\nabla_\nu\phi\partial^\mu\partial^\nu\phi\,,
\eea
where the $c_i(\da,p)$ are dimension-dependent coefficients fixed by the Kaluza-Klein reduction:
\bea
	c_2(\da,p)&=&\frac{2[n(p-3)+(p-1)^2]}{(p-1)(n+p-1)}=2(1-\da^2)\,,\\
	 c_3(\da,p)&=&\frac{2(p-1)^2-n(p^2-1)-n^2(p-3)}{4n(p-1)(n+p-1)}=\frac{3\da^2}4-\frac{p+5}{4(p-1)}+\frac{\da^{-2}}{(p-1)^2}\,,\\
	c_4(\da,p)&=&\frac{\sqrt{2}(n+1-p)}{\sqrt{n(p-1)(n+p-1)}}=\frac{2}{(p-1)\da}\l[(p-1)\da^2-1\r],\\
	c_5(\da,p)&=&\frac{n(p+n-5)-6(p-1)}{n(n+p-1)}=-\da^2+\frac{p+7}{p-1}-\frac{12}{(p-1)^2\da^2}\,.
\eea

One may check that the equations of motion for the metric \eqref{ReducedEinsteinEq} and for the scalar \eqref{ReducedPhiEq} derive from the reduced action, by replacing \eqref{ReducedEinsteingeneric} and \eqref{ReducedGBgeneric} into \eqref{GBAction} and imposing \eqref{EinsteinFrame} and \eqref{CanonicalScalar}:
\be\label{App:CurvedKKGBAction}
\begin{split}
	\bar S =& \int \ud^{p+1}x\,\sqrt{-\bar g}\l\{\bar R\l[1+2\ha \tilde Re^{-\frac{\phi}{(p-1)\da}\l(2-(p-1)\da^2\r)}\r] -2\La e^{-\da\phi}+\tilde R e^{-\frac{2\phi}{(p-1)\da}}-\frac{e^{\da\phi}}4F^{\;2}\r.\\
			&+\ha\tilde\G e^{-\frac{\phi}{(p-1)\da}(4-(p-1)\da^2)}-\half\partial\phi^2\l[1+2\ha c_5(\da,p)\tilde Re^{-\frac{\phi}{(p-1)\da}\l(2-(p-1)\da^2\r)}\r]\\
			&\left.+\ha e^{\da\phi}\l[\bar \G + c_2(\da,p)\bar G^{\mu\nu}\partial_\mu\phi\partial_\nu\phi +c_3(\da,p)\l(\partial\phi^2\r)^2+ c_4(\da,p)\partial\phi^2\square\phi  \r]\r\}\,,
\end{split}
\ee
where we have replaced the higher-dimensional gravity densities by their lower-dimensional expansions and discarded some total derivatives. This proves the consistency of the reduction: the reduced equations of motion derive from the reduced action.

%The coefficients $c_i(n,3)$ coincide with those found in \cite{Amendola:2005cr}:
%\be
%	c_2(n,3)=\frac4{2+n}\,, \qquad c_3(n,3)=\frac{1-n}{n(2+n)}\,,  \qquad c_4(n,3)=\frac{n-2}{\sqrt{n(2+n)}}\,,
%\ee
%but there is a discrepancy with $c_2(n,4)$ found in \cite{Charmousis:2003ke}:
%\be
%	c_2(n,4)=\frac{2(9+7n)}{3(n+3)}\,, \quad c_3(n,4)=-\frac{n^2+15n-18}{12n(n+3)}\,,  \quad c_4(n,4)=\sqrt{\frac{2(n-3)^2}{3n(n+3)}}\,.
%\ee
%These coefficients also agree with those of \cite{Dabrowski:2008kx}, where a generic conformal transformation is performed on the metric and the first two Lovelock invariants are calculated (in our case, this corresponds to $\ba=0$, $n=0$). They coincide with the $c_i(3,2)$ calculated in \cite{Mignemi:2007fq}, except for a minus sign in front of the first-order $\partial\phi^2$ kinetic term, and provided we set $\epsilon = \tilde R$ and not $\epsilon = k_{(n)}$ as indicated in this reference. However, we find several discrepancies with the results of \cite{Melis:2006fj,Mignemi:2006ut}, which can be compared to ours setting $p=3$, $\alpha=-n/2$ and $\beta=1$ in the formul\ae\ \eqref{ReducedEinsteingeneric} and \eqref{ReducedGBgeneric}.

\section{Products of spheres are Gauss-Bonnet spaces\label{Appendix:GBNonMaxHor}}

In this appendix, we prove that the horizon geometry for $(p+1)$-dimensional Gauss-Bonnet black holes can consist in an arbitrary number of $m$-dimensional constant curvature spaces, provided their radii are all identical.\footnote{A first version of this proof can be found in \cite{Maeda:2010bu} where equal radii are assumed from the start.}

Working in the orthogonal frame $\theta^i$, the curvature two-form of a product of $q$ $m$-dimensional constant curvature spaces $\mathbf K_1^m\times\ldots\times\mathbf K_q^m$ with normalised curvatures $k_{m,1}$, \ldots, $k_{m,q}$ can be written
\be
	\Omega^{ij} = k_ {ij}\theta^{i}\theta^j\,,\qquad k_{ij}=k_{m,r}\,\,\,(i,j)\in[rm+1,(r+1)m]^2,\,\,0\leq r\leq q.
\ee
where there is no summation in the indices. Then the first curvature scalar is
\be
	\ga_1\star\theta = \Omega^{ij}\star\theta_{ij}=\sum_{r=1}^qk_{m,r}m(m-1)\,,
\ee
while the first Lovelock tensor is
\bea
	G_i^{(1)j}\star\theta_j &=&\Omega_i^{\phantom{2}j}\star\theta_{j}\,,\\
	&=& \sum_{r=1}^qk_{m,r}\l[m(m-1)\star\theta_i-2(m-1)\da^{\phantom{2}j_r}_{i_r}\star\theta_{j_r}\r].
\eea
The Einstein space condition is expressed as
\be
	H_i^{(1)j}\star\theta_j = G_i^{(1)j}\star\theta_j-\frac{mq-2}{mq}\ga_1\star\theta_i =0\label{EinsteinSpaceCondition}
\ee
Replacing with the previous expressions, we obtain
\be
	 \sum_{r=1}^qk_{m,r}\l[2\frac{m-1}{q}\star\theta_i-2(m-1)\da^{\phantom{2}j_r}_{i_r}\star\theta_{j_r}\r]=0\,,
\ee
which projected on a given $0\leq r_0\leq q$ yields
\be
	qk_{m,r_0}= \sum_{r=1}^qk_{m,r} \qquad \forall 0\leq r\leq q
\ee
which implies that
\be
	k_{m,r}=k_{(m)} \qquad \forall 0\leq r\leq q\,.
\ee
We  now deal with the Gauss-Bonnet space condition, which states that
\be
	H_i^{(2)j}\star\theta_j = G_i^{(2)j}\star\theta_j-\frac{mq-4}{mq}\ga_2\star\theta_i\,,\label{GBSpaceCondition}
\ee
with
\bea
	\gamma_2\star\theta =  \Omega^{ij} \Omega^{kl}\star\theta_{ijkl}\,,
\eea
so that
\be
	\gamma_2=m^2(m-1)^2\sum_{r,s=1, r\neq s}^qk_{m,r}k_{m,s}+m(m-1)(m-2)(m-3)\sum_{r=1}^q(k_{m,r})^2\,,
\ee
and
\bea
	G_i^{(2)j}\star\theta_j &=&\Omega_i^{\phantom{2}j}\Omega^{kl}\star\theta_{jkl}\,,\\
				&=& (m-1)(m-2)(m-3)\sum_{r=1}^q(k_{m,r})^2\l[m\star\theta_i-4\delta_{i}^{\phantom{2}j_r}\star\theta_{j_r}\r]+\nn\\
				&&+m(m-1)^2\sum_{r,s=1, r\neq s}^qk_{m,r}k_{m,s}\l[m\star\theta_i-2\l(\delta_{i}^{\phantom{2}j_r}\star\theta_{j_r}+\delta_{i}^{\phantom{2}l_s}\star\theta_{l_s}\r)\r].
\eea
Replacing into \eqref{GBSpaceCondition}, we obtain
\bea
	0&=&4 (m-1)(m-2)(m-3)\sum_{r=1}^q(k_{m,r})^2\l[\frac{m}q\star\theta_i-4\delta_{i}^{\phantom{2}j_r}\star\theta_{j_r}\r]+\nn\\
				&&+2m(m-1)^2\sum_{r,s=1, r\neq s}^qk_{m,r}k_{m,s}\l[\frac2q\star\theta_i-\l(\delta_{i}^{\phantom{2}j_r}\star\theta_{j_r}+\delta_{i}^{\phantom{2}l_s}\star\theta_{l_s}\r)\r].
\eea
Projecting this last equation on a given index $r_0$ gives
\be
	(m-2)(m-3)\l[q(k_{m,r_0})^2-\sum_{r=1}^q(k_{m,r})^2\r]=m(m-1)\l[\sum_{r\neq s}^qk_{m,r}k_{m,s}-2qk_{m,r_0}\sum_{r\neq r_0}^qk_{m,r}\r].
\ee
Subtracting this equation for two indices $r_1$, $r_2$ finally yields
\be
	0=(k_{m,r_2}-k_{m,r_1})\l[(m-2)(m-3)(k_{m,r_2}+k_{m,r_1})+2m(m-1)\sum_{r\neq r_1,r_2}^qk_{m,r}\r]
\ee
which implies
\be
	k_{m,r}=k_{(m)}\qquad \forall 0\leq r\leq q\,.
\ee

\addcontentsline{toc}{section}{References}

\providecommand{\href}[2]{#2}\begingroup\raggedright\endgroup

\end{document}